\documentclass[final,1p,times]{elsarticle}
\pdfoutput=1 % if your are submitting a pdflatex (i.e. if you have
\usepackage{xcolor,cancel}
\usepackage{physics}
\usepackage{amssymb}
\usepackage{amsmath}

\journal{Physics of the Dark Universe}

\newcommand{\VEV}[1]{\langle #1 \rangle}
\newcommand{\eq}[1]{Eq.~(\ref{#1})}
\newcommand{\dis}[1]{\begin{equation}\begin{split}#1\end{split}\end{equation}}

\newcommand{\bfrac}[2]{\left(\frac{#1}{#2} \right)  }

\newcommand\gev{\,{\rm GeV}}

\newcommand\kev{\,{\rm keV}}
\newcommand\unitev{\,{\rm eV}}

\begin{document}

\begin{frontmatter}

\title{\boldmath Stable dark matter from Pauli blocking in the degenerate fermion background with Quantum Field Theory}

\author[a]{Wonsub Cho}
\ead{sub526@skku.edu}
\author[a,b]{Ki-Young Choi}
\ead{kiyoungchoi@skku.edu}
\author[a]{Junghoon Joh\corref{cor1}}
\ead{ttasiki1@skku.edu}
\cortext[cor1]{Corresponding author}
\author[c]{Osamu Seto}
\ead{seto@particle.sci.hokudai.ac.jp}

\affiliation[a]{Department of Physics and Institute of Basic Science, Sungkyunkwan University, 2066 Seobu-ro, Suwon-si, Gyeonggi-do, 16419, Korea}
\affiliation[b]{Korea Institute for Advanced Study, Seoul 02455, Korea}
\affiliation[c]{Department of Physics, Hokkaido University, Sapporo 060-0810, Japan}

\begin{abstract}
We study a mechanism to make dark matter stable based on the Pauli blocking in the fermion background. In the background where fermions occupy the states, the decay of dark matter to those final states is not allowed, as a result, DM becomes stable.
We derive the evolution equations of the distribution function in the quantum field theory and compare it with the Boltzmann equation.
We apply this mechanism to a realistic model of neutrino and dark matter.
\end{abstract}


\end{frontmatter}

%%%%%%%%%%%%%%
\section{Introduction}
%%%%%%%%%%%%%%
Dark matter (DM) must be stable. If it decays, the lifetime should be longer than about $170$ Gyr not to spoil the formation of large scale structures ~\cite{Ichiki:2004vi,Audren:2014bca,Nygaard:2020sow}, which is  much longer than the age of the Universe of about $14$~Gyr~\cite{Planck:2018vyg}.
%~\footnote{For multi-component DM, the sub-dominant component with a few percent to the whole DM energy density may have shorter lifetime than the age of the Universe. These fractional DM might alleviate some cosmological problems such as lensing anomaly~\cite{Chudaykin:2016yfk}.}. 
In the theoretical models of dark matter, the origin of its stability is usually assumed by an ad hoc symmetry imposed by hand or accidental symmetry from the given particle contents and gauge symmetry~\cite{Hambye:2010zb}. Otherwise, the mass of DM needs to be small and/or the interaction coupling constants are very small so that the decay rate becomes very small, such as for axions or sterile neutrinos~\cite{Baer:2014eja}.

Another possibility is to use the Pauli blocking in the DM decay to fermions~\cite{Bjaelde:2010vt,Batell:2024hzo}.
Bjaelde and Das studied the decay of DM into the Fermi sea of neutrinos, where the Pauli blocking controls the DM decay~\cite{Bjaelde:2010vt} which is analogous to the situation of fermion production by preheating~\cite{Greene:1998nh,Greene:2000ew}. They considered Yukawa couplings of the order of the unity and applied the change of the equation of state of DM at the last stage to find solutions of the cosmological problems. 
Ghosh and Mukhopadhyay studied DM momentum distribution from the decay of an inflaton, where the Pauli blocking modifies the momentum distribution of DM and its energy density~\cite{Ghosh:2022hen}.
Pauli blocking also affects the pair creation of fermion fields in strong electric field~\cite{Prakapenia:2023tsw}.

Recently, Batell and Yin have considered the cosmic stability of DM from the Pauli blocking, including the effect of scatterings and parametric resonances as well as the decay, and explored applications to astrophysics and cosmology~\cite{Batell:2024hzo}.

In this paper, we study the evolution of the DM decaying into fermions in the quantum field theory (QFT) as well as in the formalism of Boltzmann equation. We find several approximate solutions and compare them with the numerical results. Finally, we suggest a realistic model of particle physics with the neutrinos, where the lifetime of DM becomes longer due to the Pauli blocking of neutrino background.

In Sec.~\ref{sec:BE}, we show the evolution of DM and fermions in the formalism of the Boltzmann equation with analytic and numerical results in the non-expanding and expanding Universe. In Sec.~\ref{sec:QFT} we derive the evolution equations in the QFT formalism and show the numerical results with some semi-analytical calculation. In Sec.~\ref{sec:Model}, we show a specific example where the lifetime of DM is extended and make the model viable. In Sec.~\ref{sec:con}, we conclude this paper.

%%%%%%%%%%%%%%%%%
\section{Particle Production with Boltzmann Equation}
\label{sec:BE}
%%%%%%%%%%%%%%%%%
We consider a real scalar $\phi$ which can decay only into fermion pairs $\psi$ and $\bar{\psi}$.
As a simple example, we assume the following Yukawa interaction  with the coupling $\lambda$ in Lagrangian~\footnote{We do not consider scattering processes, since those are subdominant for small Yukawa coupling as we consider in this paper.}
\dis{
{\mathcal L} \ni \lambda \bar{\psi}\psi \phi.
\label{Yukawa}
}

For two body decay, the non-relativistic scalar $\phi$ with negligible momentum will produce two fermions with the same energy $E=m_\phi/2$ with opposite directions,
\dis{
\phi (p)\rightarrow \psi(k) + \bar{\psi}(k'),
}
where $p=(m_\phi,{\bf 0})$, $k=(E,\vec{\bf k})$, and $k'=(E,-\vec{\bf k})$,
with $|\vec{\bf k}| = p_\psi \equiv\frac{m_\phi}{2} \sqrt{1-R^2}$, $E^2=m_\psi^2+|\vec{\bf k}|^2 $, and  $R \equiv 2m_\psi / m_\phi$.
The decay rate in the vacuum is given by
\dis{
\Gamma_{\phi} =\frac{p_\psi}{8\pi m_\phi^2}|\mathcal{M}|^2  = \frac{\lambda^2 m_\phi}{8\pi}\left(1- R^2 \right)^{3/2},
}
and the corresponding lifetime is
\dis{
\tau_{\phi,0}=\frac{1}{\Gamma_{\phi}} \simeq 1.6 \times 10^{13} \sec \bfrac{10^{-12}}{\lambda}^2\bfrac{10^{-3}\unitev}{m_\phi} \left(1-R^2 \right)^{-3/2}.
\label{lifetime0}
}
Even for a small Yukawa coupling and a light scalar field, the lifetime in the vacuum is supposed to be much shorter than the age of the Universe, and it seems impossible for the scalar $\phi$ to be a candidate of DM. However, as was pointed out in Ref.~\cite{Bjaelde:2010vt} and we will see below, if we consider the statistics of the fermions and the back-reaction, the decay of $\phi$ is delayed due to the Pauli blocking in the fermion background and its lifetime can be much longer than the age of the Universe. 

The explicit Boltzmann equation for the scalar phase space distribution $f_\phi(\vec{\bf p},t)$ is given by
\dis{
\frac{\partial f_\phi(\vec{\bf p},t)}{\partial t} - H p \frac{\partial f_\phi(\vec{\bf p},t)}{\partial p}=f_\phi^{\rm coll}(\vec{\bf p},t),
\label{be_fphi}
}
where the right-hand side is the collision term for $\phi$ and $H$ is the cosmic expansion rate.
For the process $\phi(\vec{\bf p})\leftrightarrow \psi(\vec{\bf k}) +\bar{\psi}(\vec{\bf k'})$, the collision term becomes
\dis{
f_\phi^{\rm coll}(\vec{\bf p},t) &= \frac{1}{2E_{\vec{\bf p}}}\int \frac{d^3\vec{\bf k'} }{(2\pi)^32E_{\vec{\bf k'}}}\frac{d^3\vec{\bf k }}{(2\pi)^32E_{\vec{\bf k}}} (2\pi)^4\delta^4(p-k'-k)|\mathcal{M}|^2 \\
&\times\left[ f_\psi(\vec{\bf k},t)f_{\bar{\psi}}(\vec{\bf k}',t)(1+f_\phi(\vec{\bf p},t)) -f_\phi(\vec{\bf p},t)(1-f_\psi(\vec{\bf k},t))(1-f_{\bar{\psi}}(\vec{\bf k'},t))\right],
}
with
\dis{
|\mathcal{M}|^2 = 2\lambda^2m_\phi^2 \left(1- R^2 \right).
}
The number density of $\phi$ is given by
\dis{
n_\phi(t) = \int \frac{d^3\vec{\bf p}}{(2\pi)^3} f_\phi(\vec{\bf p},t).
}  In the case when $f_\phi(\vec{\bf p},t) \gg 1$, and $\phi$ is at rest, by integrating the Boltzmann \eq{be_fphi}, we obtain its evolution equation
\dis{
\dot{n}_\phi(t) +3H n_\phi(t) &\simeq  - \Gamma_\phi  n_\phi(t)\left.\left[ 1-2 f_\psi(\vec{\bf k},t) \right]\right|_{|\vec{\bf k}|=p_\psi},
}
where we imposed isotropy $f_\psi({\vec{k}},t)=f_\psi({|\vec{k}|},t)$ and CP symmetry with $f_\psi({\vec{k}},t)=f_{\bar{\psi}}(\vec{k},t)$.
Here, we can see explicitly that the decay of $\phi$ is blocked when $f_\psi$ is larger than $1/2$, and the inverse decay occurs.

The Boltzmann equation of the phase distribution function of the fermion for one degree of freedom (a single spin), $f_\psi(\vec{\bf p},t)$, is given by
\dis{
\frac{\partial f_\psi(\vec{\bf k},t)}{\partial t} - H k \frac{\partial f_\psi(\vec{\bf k},t)}{\partial k}= \frac{1}{2}f_\psi^{\rm coll}(\vec{\bf k},t).
\label{be_fpsi}
}
For the process $\phi(\vec{\bf p})\leftrightarrow \psi(\vec{\bf k}) +\bar{\psi}(\vec{\bf k'})$, the collision term becomes
\dis{
f_\psi^{\rm coll}(\vec{\bf k},t) &= -\frac{1}{2E_{\vec{\bf k}}}\int \frac{d^3\vec{\bf k'} }{(2\pi)^32E_{\vec{\bf k'}}}\frac{d^3\vec{\bf p }}{(2\pi)^32E_{\vec{\bf p}}} (2\pi)^4\delta^4(p-k'-k)|\mathcal{M}|^2 \\
&\times\left[ f_\psi(\vec{\bf k},t)f_{\bar{\psi}}(\vec{\bf k'},t)(1+f_\phi(\vec{\bf p},t)) -f_\phi(\vec{\bf p},t)(1-f_\psi(\vec{\bf k},t))(1-f_{\bar{\psi}}(\vec{\bf k'},t))\right],
}
and in the case when $f_\phi(\vec{\bf p},t) \gg 1$, and $\phi$ is at rest, it can be written as 
\dis{
f_\psi^{\rm coll}(\vec{\bf k},t) & \simeq\Gamma_\psi n_\phi(t)\bfrac{2\pi^2}{p_\psi^2}\delta(|\vec{\bf k}|-p_\psi)\left[ (1-f_\psi(\vec{\bf k},t))(1-f_{\bar{\psi}}(-\vec{\bf k},t))  - f_\psi(\vec{\bf k},t)f_{\bar{\psi}}(-\vec{\bf k},t) \right] \\
&= \Gamma_\psi n_\phi(t)\bfrac{2\pi^2}{p_\psi^2}\delta(|\vec{\bf k}|-p_\psi)\left[ 1-2 f_\psi(\vec{\bf k},t) \right].
} 
%

%%%%%%%%%%%%%%%%%
\subsection{Without Cosmic Expansion}
\label{sec:NoExp}
%%%%%%%%%%%%%%%%%
In this section, we consider non-expanding Universe and give the analytic solution of the Boltzmann equation. The Boltzmann equations for $f_\psi$ and $n_\phi$ are
\begin{equation}
\frac{\partial f_\psi(\vec{\bf k},t)}{\partial t} = \Gamma_\phi n_\phi(t) \bfrac{\pi^2}{p_\psi^2}\frac{1}{p_\psi}\delta ( |\vec{\bf k}|/p_\psi -1 ) \left[ 1-2 f_\psi(\vec{\bf k},t) \right],
\label{BE_noxpand}
\end{equation}
and
\dis{
\dot{n}_\phi(t) &\simeq  - \Gamma_\psi  n_\phi(t)\left.\left[ 1-2 f_\psi(\vec{\bf k},t) \right]\right|_{|\vec{\bf k}|=p_\psi}.
\label{nphieq}
}
If we integrate the sum of Eq.~(\ref{nphieq}) and Eq.~(\ref{BE_noxpand}) times $\frac{p_\psi^3}{\pi^2}$, 
we obtain the equation of total number density conservation given by
\dis{
n_\phi(t)+\frac{p_\psi^3}{\pi^2} \left.f_\psi(\vec{\bf k},t)\right|_{|\vec{\bf k}|=p_\psi }
=  n_\phi(t_i) + \frac{p_\psi^3}{\pi^2} \left.f_\psi(\vec{\bf k},t_i)\right|_{|\vec{\bf k}|=p_\psi } \equiv C_0.
 \label{nfconst}
}
Using \eq{nfconst}, we can replace $f_\psi$ in \eq{nphieq} by $n_\phi$ to obtain an differential equation for $n_\phi(t)$
\dis{
\dot{n}_\phi(t) &\simeq  - \Gamma_\psi  n_\phi(t)\left.\left[ 1-2 \{C_0 - n_\phi(t)\}\frac{\pi^2}{p_\psi^3} \right]\right|_{|\vec{\bf k}|=p_\psi} \\
&=-A n_\phi(t) +Bn_\phi^2(t),
\label{nphieq2}
}
with
\begin{align}
A \equiv  \Gamma_\psi \left(1-C_0 \frac{2\pi^2}{p_\psi^3} \right),  \qquad \mathrm{and} \qquad 
B \equiv  - \Gamma_\psi \frac{2\pi^2}{p_\psi^3} .
\end{align}
This equation is a type of Bernoulli equation and can be solved exactly, since the equation for $1/n(\phi)$ is a first-order linear differential equation. The solution is 
\dis{
\frac{1}{n _\phi(t)}& = \frac{B}{A} +\left( \frac{1}{n_\phi(t_i)} -\frac{B}{A}\right) \exp[A(t-t_i)] \\
= & \left[ C_0 - \frac{p_\psi^3}{2\pi^2}   \right]^{-1} 
+ \left( \frac{1}{n _\phi(t_i)}  - \left[ C_0 - \frac{p_\psi^3}{2\pi^2}   \right]^{-1}  \right)
\exp\left[\Gamma_\psi\left( 1-C_0 \frac{2\pi^2}{p_\psi^3}  \right) (t-t_i)\right],
\label{sol_static}
} 
with $t_i$ being an initial time.
\begin{figure}
    \centering
    \includegraphics[width=0.55\textwidth]{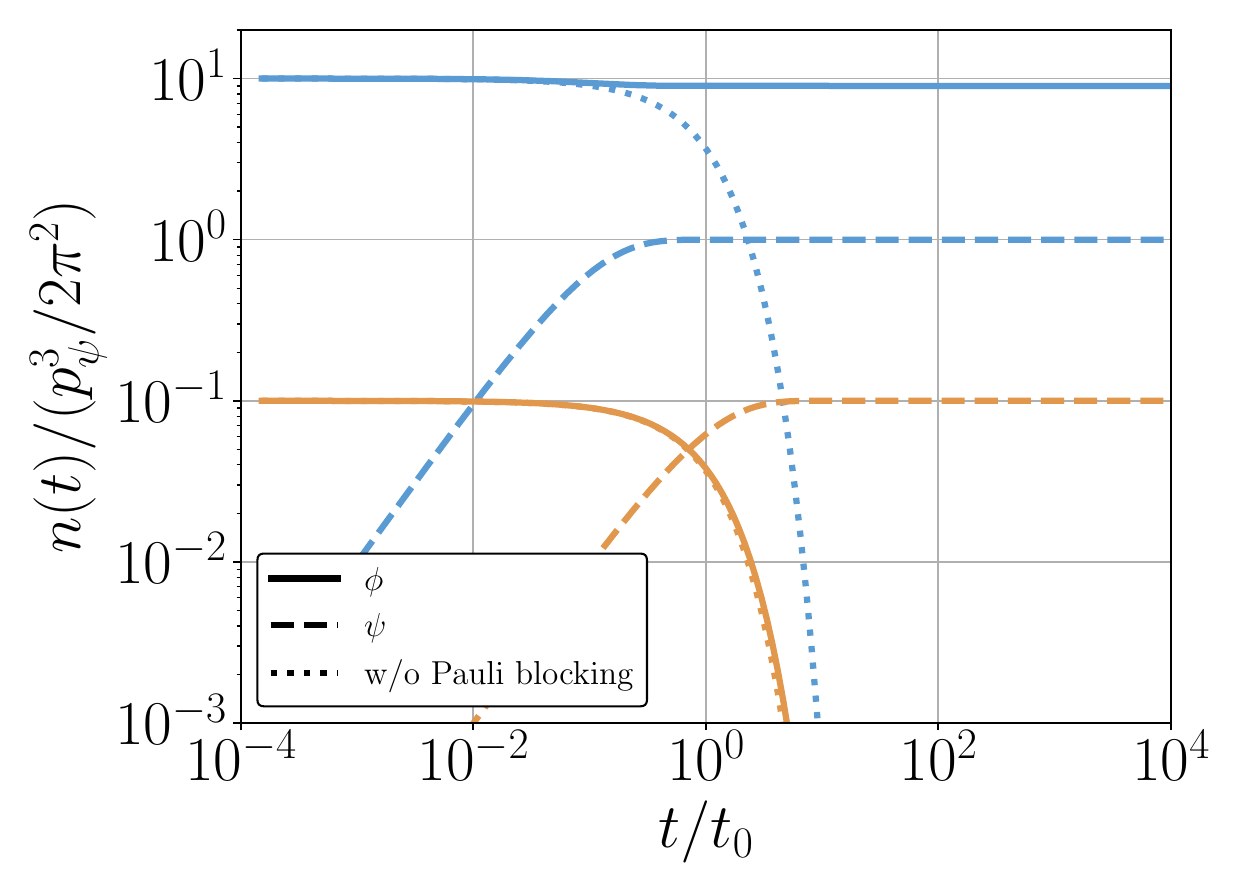}
    \caption{ Evolution of the number density of $\phi$ (solid) and $\psi$ (dashed)  normalized by $p_\psi^3/2\pi^2$ for $C_0/(\frac{p_\psi^3}{2\pi^2}) = 0.1,$  (orange) and $10$  (blue), respectively. Here, we used $m_\phi = 10^{-3} \unitev, \ f_\psi(t_i)=0$ and $\Gamma_\phi=10^{-37}\gev$, with $t_0 = 1/\Gamma_\phi$.
  }
    \label{fig:BE_nonexpanding}
\end{figure}

There are two different physical cases:
For Case (I), the initial number density of $n_\phi$ is not enough to fill the phase space of $\psi$ from its decay, so $\phi$ can decay until it depletes its density. For Case (II), the initial number density of $n_\phi$ is large enough to fill the phase space of $\psi$ and there still remains some amount after filling the Fermi phase space. For each case,
\dis{
(I) \quad C_0 <  \frac{p_\psi^3}{2\pi^2}:& \quad n_\phi(t) \rightarrow 0, \quad f_\psi \rightarrow\frac{\pi^2}{p_\psi^3} C_0, \\
& \quad n_\psi(t) \rightarrow g\int\frac{d^3k}{(2\pi)^3}\frac{\pi^2}{p_\psi^3} C_0 \delta(|\vec{\bf k}|/p_\psi -1) =C_0,\\
(II)\quad  C_0 >  \frac{p_\psi^3}{2\pi^2}:& \quad n_\phi(t) \rightarrow C_0- \frac{p_\psi^3}{2\pi^2} = n_\phi(t_i) - \frac{p_\psi^3}{\pi^2} \left( \frac12- \left.f_\psi(\vec{\bf k},t_i)\right|_{|\vec{\bf k}|=p_\psi }  \right),\\
& \quad f_\psi \rightarrow \frac12, \quad n_\psi(t) \rightarrow g\int\frac{d^3k}{(2\pi)^3}\frac12 \delta(|\vec{\bf k}|/p_\psi -1)  = \frac{p_\psi^3}{2\pi^2},
\label{BEsol1}
}
where we used $g=2$ for two spins of the fermion.

In Fig.~\ref{fig:BE_nonexpanding}, we show the evolution of the number density for $\phi$ and $\psi$ normalized by $p_\psi^3/2\pi^2$, for  both case with  (I) $C_0 < p_\psi^3/2\pi^2$ (orange) and (II) $C_0 > p_\psi^3/2\pi^2$ (blue). For case (I), $\phi$ completely decays to fermion,  while  for case (II) only small fraction of $\phi$ decays to fermion whose phase space is fully occupied.

For a scalar field which is initially highly degenerate  with $C_0\simeq n_\phi(t_i) \gg p_\psi^3/2\pi^2$, we can approximate the solution as
\dis{
n_\psi(t) &= n_{\bar{\psi}}(t) \simeq \frac{p_\psi^3}{2 \pi^2}\left( 1-\exp[-n_\phi(t_i)\Gamma_\phi \frac{\pi^2}{p_\psi^3}(t-t_i)] \right),\\
n_\phi(t) &\simeq n_\phi(t_i)-n_\psi(t).
}
Initially, $n_\psi$ grows linearly with time as $n_\psi \sim n_\phi(t_i)\Gamma_\phi t$ and later fills the Fermi surface nearly at the time scale
\dis{
t_c\simeq \bfrac{n_\phi(t_i)}{p_\psi^3/\pi^2}^{-1}\Gamma_\phi^{-1},
\label{tc_BE_nonexpand}
}
which is smaller than the lifetime of scalar when there is no Pauli blocking, $\Gamma_\phi^{-1}$.
%compared to $\Gamma_\phi^{-1}$ when there is no Pauli blocking.

Note that the scatterings $\phi\phi \leftrightarrow \psi\psi$ and $\phi\psi \leftrightarrow \phi\psi$ can happen in the order of $\lambda^4$ and does not affect our results once the timescale of the scattering is much smaller than the Hubble time. However, in the opposite case when the scattering time-scale becomes shorter than the expansion, the scatterings can redistribute the momentum of $\phi$ and $\psi$, which can initiate the decrease of condensate $\phi$~\cite{Batell:2024hzo}.

%%%%%%%%%%%%%%%%%
\subsection{With Cosmic Expansion}
\label{sec:Exp}
%%%%%%%%%%%%%%%%%
In the expanding Universe, the magnitude of a physical momentum $k(t)\equiv|\vec{\bf k}(t)|$   redshifts inversely proportional to the scale factor in the metric of the spacetime. That is also true for the number density of the scalar field, in this case $n_{\phi} \propto a^{-3}$. Under these changes of the background, the scalar field $\phi$ can decay and fill the Fermi surface. However, the redshift of the momentum makes the Fermi surface empty and allows the decay of $\phi$. This continues until the number density of $\phi$ becomes smaller than the number density in the Fermi surface of momentum $p_\psi$, when Pauli blocking is not effective any more.

We can write the redshift of the physical momentum with the scale factor as
\dis{
k(t_p, t) = \frac{a(t_p)}{a(t)} p_\psi,
} 
where $t_p$ is the production time of the fermion $\psi$ with momentum $p_\psi$.
In matter-dominated Universe with $a(t)^3 \propto t^2$, $k(t)$ has the relation to the time as 
\dis{
t_p = t\bfrac{|\vec{\bf k}(t)|}{p_\psi}^{3/2}.
}
For a given comoving momentum, we can write 
the Boltzmann \eq{be_fpsi} in terms of a total derivative with respet to time as~\cite{Ghosh:2022hen}
\dis{
\frac{ df_\psi(t_p,t)}{d t} =f_\psi^{\rm coll}(t_p,t)\simeq  \Gamma_\phi n_\phi(t)\bfrac{\pi^2}{p_\psi^2} \frac{\delta(t-t_p)}{p_\psi H(t_p)}\left[ 1-2 f_\psi(\vec{\bf k}(t),t) \right],
\label{be2_fpsi}
}
where we used  
\dis{
\delta(k(t)-p_\psi) = \frac{\delta(t-t_p)}{p_\psi H(t_p)}.
}
The analytic solution for this is
\dis{
f_\psi(t_p,t) =\frac12\left(  1-e^{-\bar{f}(t_p)}\right)\theta(p_\psi - k(t)),
}
with 
\dis{
\bar{f}(t_p) = \frac{\pi^2\Gamma_\phi }{p_\psi^3}\frac{n_\phi(t_p)}{H(t_p)}.
\label{eq:be_psi}
}
The Fermi surface of the momentum $k=p_\psi$, i.e. $f_\psi=1/2$, keeps being filled when the condition $\bar{f}(t_p)\gg 1$ is still valid.

\begin{figure*}[!t]
\begin{center}
\begin{tabular}{cc} 
 \includegraphics[width=0.45\textwidth]{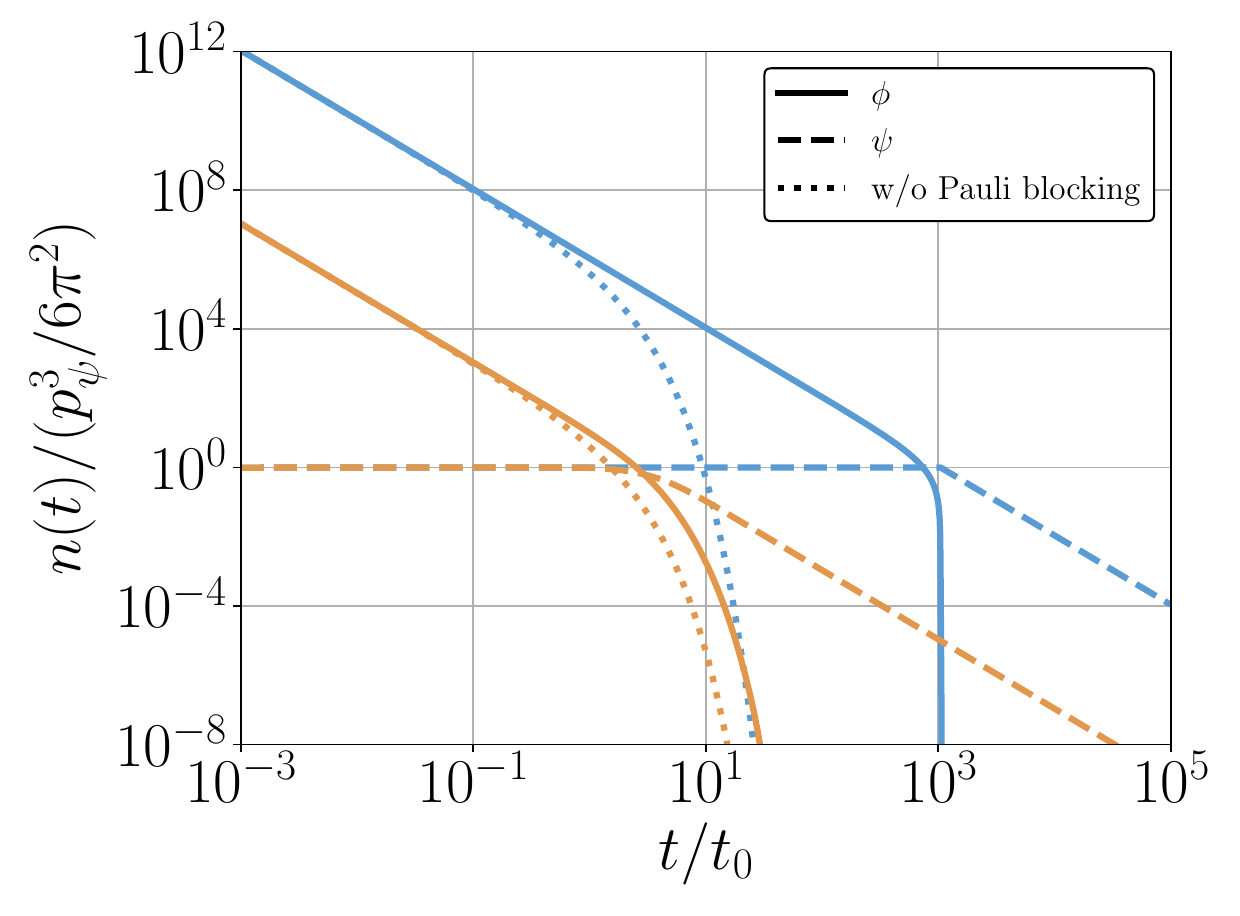}
 &
  \includegraphics[width=0.45\textwidth]{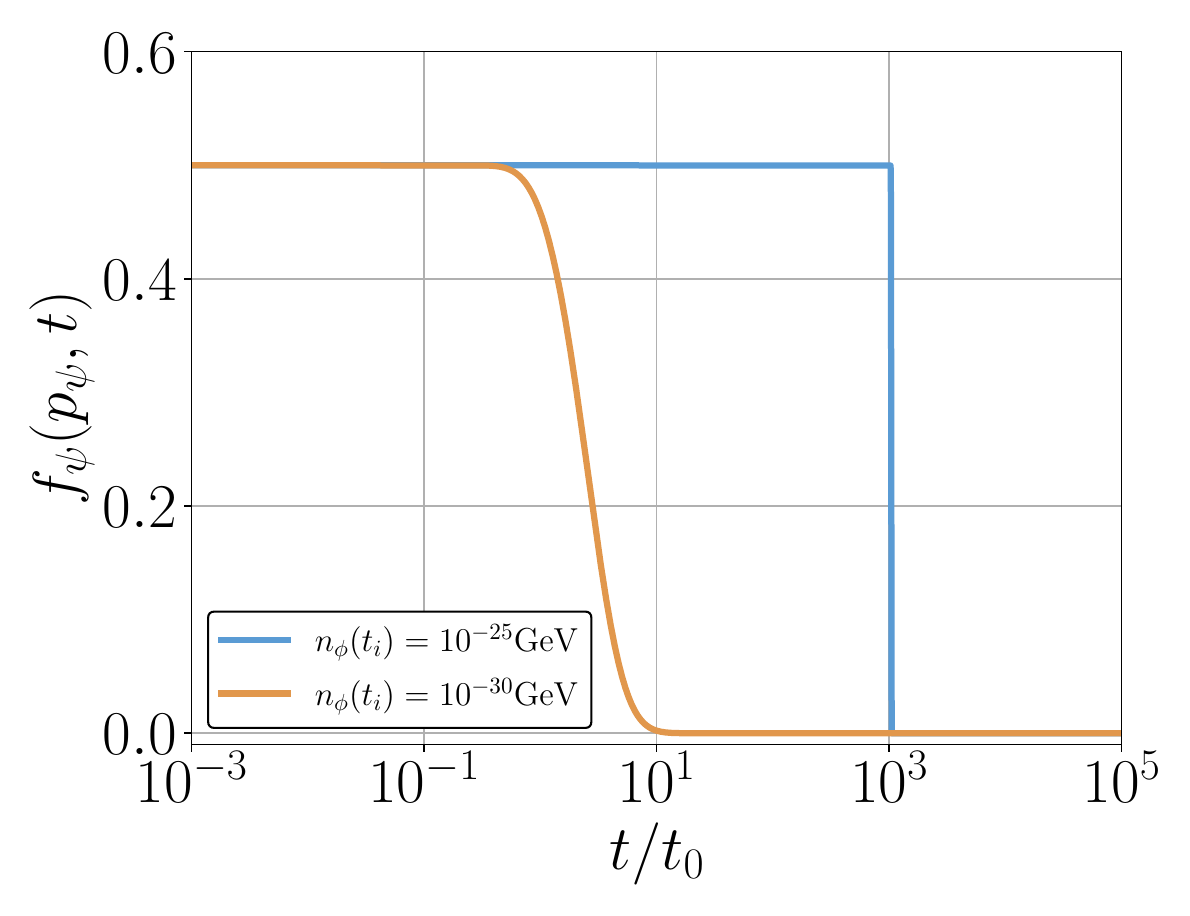}
  \end{tabular}
 \caption{The time evolution of the number density of scalar (solid) and fermion (dashed) (left) and the distribution function of the produced fermions at momentum $p_\psi$ (right). Here, we used $m_\phi = 10^{-3} \unitev, \ f_\psi (p_\psi,t_i)=0$ and $\Gamma_\phi=10^{-37}\gev$, with $t_0 = 1/\Gamma_\phi$. The orange and blue lines correspond to the different initial number density, $n_\phi(t_i) = 10^{-30} ,10^{-25} \, \mathrm{GeV}^3$, respectively. The lifetime becomes longer for larger number density due to the Pauli blocking.}
 \label{fig:BE_expanding}
\end{center}
\end{figure*}

In Fig.~\ref{fig:BE_expanding} (left), we show the time evolution  of the number density of scalar dark matter (solid line) and fermion field (dashed line) for different initial number density of scalar field
 with $10^{-30}\, \mathrm{GeV}$ (orange) and $n_\phi(t_i) = 10^{-25}\, \mathrm{GeV}$ (blue), respectively. 
For comparison, we also show the case of scalar decay without Pauli blocking (dotted line) for each case.
When the number density of $\phi$ is large enough compared to that in the Fermi sphere,
\dis{
n_\phi(t) \gg \frac{p_\psi^3}{6\pi^2},
\label{nphibig}
}
the fermions produced from the decay fills the fermi surface $k=p_\psi$ quickly with $f_k=1/2$  and the number density of them is 
\dis{
n_\psi(t)=&g\int^{p_\psi} \frac{d^3k}{(2\pi)^3} f_\psi (k) = \frac{p_\psi^3}{6\pi^2},
}
which is constant with time. At this stage,
most of the fermions are relativistic and its energy density is also constant as
\dis{
\rho_\psi = \frac{p_\psi^4}{8\pi^2}.
}
From the conservation of energy in the expanding Universe, we can show that
\dis{
\frac{d}{dt}(\rho_\phi + \rho_\psi) + 3H\rho_\phi +4H\rho_\psi=0.
}
 Therefore, the energy density of $\phi$ follows~\cite{Bjaelde:2010vt}
\dis{
\rho_\phi(t) = -\frac{p_\psi^4}{6\pi^2} + \left(\rho_\phi(t_i) + \frac{p_\psi^4}{6\pi^2}\right)\bfrac{a(t)}{a(t_i)}^{-3},
\label{rhophibig}
}which is valid when $n_\phi(t) \gg p_\psi^3/6\pi^2$.

As the Universe expands, the number density of $\phi$ becomes small. When
\dis{
n_\phi(t) \lesssim \frac{p_\psi^3}{6\pi^2},
\label{nphi_t}
}
the number density of $\phi$ is not enough to fill the Fermi sphere and $\phi$ rapidly decays and $n_\psi$ begins to decrease only due to expansion of the Universe.
Therefore, we can estimate the extended lifetime from the equality in \eq{nphi_t}, $n_\phi(t)= \frac{p_\psi^3}{6\pi^2}$.
Using $n_\phi(t) \simeq n_\phi(t_{i})(a(t)/a_{i})^{-3}$, and $a(t)\propto t^{2/3}$ in matter-dominated era, we find that
\dis{
\tau_\phi \simeq  t_{i}\bfrac{\rho_{\phi}(t_i)}{m_\phi}^{1/2}\bfrac{p_\psi^3}{4\pi^2}^{-1/2}.
}

In Fig.~\ref{fig:BE_expanding} (right), we show the time evolution of the momentum distribution function of fermions at momentum $k=p_\psi$.
Since we assume that the scalar decays at rest frame, the momentum of a produced fermion is fixed as $p_\psi$ right after it is produced. The change of momentum comes only from the redshift, $k(t)=p_\psi a(t_p)/a(t)$. Therefore, the momentum distribution function as a function of momentum at a given time $f_\psi (k, t_0)$ has a scaling  relation with $f_\psi$ at $p_\psi$ at a time $t_0 \bfrac{k}{p_\psi}^{3/2}$ as
\dis{
f_\psi (k, t_0) = f_\psi \left(p_\psi, t_0 \bfrac{k}{p_\psi}^{3/2}\right).
}

%%%%%%%%%%%%%%%%%
\section{Particle Production with the Quantum Field Theory}
\label{sec:QFT}
%%%%%%%%%%%%%%%%%

In this section, we consider quantum field theory (QFT) for the fermion production from oscillating scalar field. Scalar field production from oscillating scalar field based on QFT was studied in Ref.~\cite{Moroi:2020bkq}, and they compared the results with those from Boltzmann equations. Here, we follow them to derive the equations for fermion-antifermion pair production from oscillating scalar field.
First, we apply them to the non-expanding Universe and then  consider expansion.

We decompose fermion and antifermion field $\psi$ and $\bar{\psi}$ by using the creation and annihilation operators,
\dis{
\psi(x) &=\sum_{s} \int \frac{d^{3} p}{(2 \pi)^{3}} \frac{1}{\sqrt{2 \omega_{p}}}\left(a_{\vec{p}}^{s}(t) u_{\vec{p}}^{s} e^{-i \vec{p} \cdot \vec{x}}+b_{\vec{p}}^{s \dagger}(t) v_{\vec{p}}^{s} e^{i \vec{p} \cdot \vec{x}}\right), \\
\bar{\psi}(x) &=\sum_{s} \int \frac{d^{3} p}{(2 \pi)^{3}} \frac{1}{\sqrt{2 \omega_p}}\left(a_{\vec{p}}^{s \dagger}(t) \bar{u}_{\vec{p}}^{s} e^{i \vec{p} \cdot \vec{x}}+b_{\vec{p}}^{s} (t)\bar{v}_{\vec{p}}^{s} e^{-i \vec{p} \cdot \vec{x}}\right),
}
where $\omega_p=\sqrt{m_\psi^2+|\vec{p} |^2 }$ and the creation and annihilation operators satisfy the anti-commutation relations as
\dis{
\{a^r_{ \vec{\bf k}}, a^{s\dagger}_{ \vec{\bf p}} \}= \{b^r_{\vec{\bf k}}, b^{s\dagger}_{ \vec{\bf p}} \}=(2\pi)^3\delta^{(3)}({ \vec{\bf k}}-{ \vec{\bf p}}) \delta^{rs},
}
with others equal to zero.
The free Hamiltonian is 
\dis{
H_{\rm free} =  \int \frac{d^{3} p}{(2 \pi)^{3}} \sum_{s} \omega_{ \vec{\bf p}} ( a^{s\dagger}_{ \vec{\bf p}} a^{s}_{ \vec{\bf p}}  +  b^{s\dagger}_{\vec{\bf p}} b^{s}_{\vec{\bf p}}).
}

The expectation value of the operator $\mathcal{O}$ is defined by 
\dis{
\expval{\mathcal{O}} \equiv {\rm Tr}(\rho \mathcal{O}), 
\label{eq:def:Oexp}
}
using a density matrix $\rho$ which evolves as
\begin{equation}
\rho(t) = U(t,t_0) \rho(t_0) U^{\dagger}(t,t_0),
\label{eq:UrhoU}
\end{equation}
where $U(t,t_0)$ is an unitary matrix. The evolution equation is given by 
\begin{equation}
i\frac{d}{dt}\rho(t) = \left[ H, \rho\right] ,
\label{eq:eom:rho}
\end{equation}
 with $H$ being the total Hamiltonian, which can be divided into the free part and the interaction part as $H = H_\mathrm{free} +H_\mathrm{int}$.
By substituting Eq.~(\ref{eq:eom:rho}) into Eq.~(\ref{eq:def:Oexp}), we obtain 
\begin{align}
    \frac{d}{dt}\expval{\mathcal{O}} =& -i \expval{\comm{\mathcal{O}}{H_{\rm int}}} + \dot{\expval{\mathcal{O}}},
\label{tderiavg} \\
\dot{\mathcal{O}} =& -i[\mathcal{O}, H_{\rm free}].
\label{tderi}
\end{align}
Applying Eq.~(\ref{tderiavg}) with the interaction Hamiltonian (\ref{Yukawa}) to the operators of fermions
$a_{\vec{p}}^{\dagger} a_{\vec{p}}$, $ b_{\vec{p}}^{\dagger} b_{\vec{p}}$, and $ b_{\vec{p}} a_{\vec{p}} $,
we obtain differential equations 
\begin{equation}
\begin{split}
    \dot{f}_{\psi,\vec{k}} & = i \lambda \phi_c(t) \left(g_{\vec{k}} - g^*_{\vec{k}} \right),\\
    \dot{f}_{\bar{\psi},\vec{k}} & = i \lambda \phi_c(t) \left(g_{-\vec{k}} - g^*_{-\vec{k}} \right),\\
    \dot{g}_{\vec{k}} & =-i\left[2 \omega_{k} g_{\vec{k}}(t)+\lambda \phi_c(t)\left\{ \frac{k^2}{\omega_k^2} \left(1-f_{\psi, \vec{k}}-f_{\bar{\psi}, -\vec{k}}\right)+\frac{2m_\psi}{\omega_k} g_{\vec{k}}\right\}  \right],
    \label{fg}
\end{split}
\end{equation}
where $f_{\vec{k}}$ and $g_{\vec{k}}$ are defined by
\begin{subequations}
\begin{align}
f_{\psi,\vec{k}}(t) &\equiv \frac{1}{2V}\sum_{r} \left\langle a_{\vec{k}}^{r \dagger} a_{\vec{k}}^{r}\right\rangle, \\
f_{\bar{\psi},\vec{k}}(t) &\equiv  \frac{1}{2V}\sum_{r} \left\langle b_{\vec{k}}^{r \dagger} b_{\vec{k}}^{r}\right\rangle, \\
g_{\vec{k}}(t) &\equiv \frac{1}{4 V \omega_{k}} \sum_{r} \sum_{s}\left\langle b_{-\vec{k}}^{s} a_{\vec{k}}^{r} \right\rangle \bar{v}_{-\vec{k}}^{s} u_{\vec{k}}^{r}, \\
g_{\vec{k}}^{*}(t) &\equiv\frac{1}{4 V \omega_{k}} \sum_{r} \sum_{s}\left\langle a_{\vec{k}}^{r \dagger} b_{-\vec{k}}^{s \dagger} \right\rangle \bar{u}_{\vec{k}}^{r} v_{-\vec{k}}^{s} .
\end{align}
\end{subequations}
We note that the number density operator $n_{\vec{p}}$ in the density matrix formalism is related in terms of the distribution function $f_{\vec{p}}$ as
\begin{equation}
n =  \int\frac{d^3p}{(2\pi)^3}f_{\vec{p}} .
\end{equation} 

Since we consider that fermions are produced only from the decay of scalar field, initial quantum states are taken to be the vacuum state $|0\rangle$ so that the density operator is $\rho(0)=|0\rangle\langle0|$ which gives the initial condition 
\dis{
f_{\psi,\vec{k}}(0) = f_{\bar{\psi},\vec{k}}(0) = g_{\vec{k}} (0) =0.
}
Since the distribution of fermion and anti-fermion are isotropic, they are independent on the direction of momentum, so that $f_{\psi,\vec{k}}(t) = f_{\bar{\psi},\vec{k}}(t) = f_{k} (t) $, and $g_{\vec{k}} (t) = g_k (t)$ with $k=|\vec{k}|$.

%The interaction Hamiltonian for Yukawa interaction \eq{Yukawa} is
%\dis{
%H_{\rm int}&(t) = \lambda \int d^3x \phi(x)\overline{\psi} (x) \psi(x)  \\
%\simeq & \lambda \phi_{\cred{c}}(t) \int \frac{d^{3} p}{(2 \pi)^{3}}\sum_{s,s'}\frac{1}{2\omega_p}\left[a^{s\dagger}_p a^{s'}_p \bar{u}^s_p u^{s'}_p + b^{s\dagger}_p b^{s'}_p \bar{v}^s_p v^{s'}_p + b^{s}_{p}a^{s'}_{-p} \bar{v}^{s}_{p}u^{s'}_{-p} + a^{s\dagger}_{p}b^{s'\dagger}_{-p} \bar{u}^{s}_{p}v^{s'}_{-p} \right].
%\label{Hint}
%}

In principle, we could also consider the distribution function of the scalar field in terms of the creation and annihilation operators and derive the evolution equation, which enables us to follow the number density of the scalar field when it starts to decay. However, in this study we focus on the stability of the scalar particle and only show that the production of the fermions are restricted due to the Pauli blocking. The full treatment might be beyond the present study and will be done in another publication.

In the following,
we use an approximation that the scalar field is dominated by the classical field $\phi(x)\simeq \phi_c(t) $.
The quantum particles of the scalar field might be produced from the back-reaction of  $\psi \bar{\psi} \to \phi$. However, since we will focus only on the timescales where
the number density of the fermions are negligible compared to that of the classical scalar field, the number density of the quantum scalar particle also can be ignored in our consideration. 
%\dis{
%H_{\text{int}}= \lambda \int d^3x \left( \phi_c(x) + \phi_q (x) \right) \overline{\psi} (x) \psi(x)
%}

%\cred{We assumed that the decaying scalar field is homogeneous and classical.

%\cred{
%The scalar field $\phi$ which has interaction with fermion and antifermion field is separated by the background classical field $\phi_c$ and quantum field $\phi_q$ produced by back-reaction or scattering. The scalar field is decomposed by the summation of the background classical field and the quantum field,
%\dis{
%\phi (x) = \phi_c (x) + \phi_q (x) = \bar{\phi} \cos m_\phi t + \int \frac{d^3q}{(2 \pi)^3} \frac{1}{\sqrt{2 \omega_q}} (c_{\vec{q}}  \, e^{i \vec{q} \cdot \vec{x}} + c_{\vec{q}}^\dagger \, e^{-i \vec{q} \cdot \vec{x}}),
%}
%where $c_{\vec{q}}$ and $c_{\vec{q}^\dagger}$ is annihilation operator and creation operator respectively.
%}

%

%%%%%%%%%%%%%%%%%
\subsection{Without Cosmic Expansion}
\label{sec:QNoExp}
%%%%%%%%%%%%%%%%%
In the non-expanding case, the amplitude of the oscillating scalar field $\bar{\phi}$ is constant and we can write it as
\dis{
\phi_c(t) = \bar{\phi}\cos m_\phi t.
\label{phidef}
}
If we rewrite the complex function $g_k$ in terms of real functions $\xi_k(t)$ and $\eta_k(t)$ as
\dis{
g_k (t) = e^{-2 i \omega_k t} [\xi_k (t) + i \eta_k (t)],
}
and introducing dimensionless variables,
\dis{
\tau = m_\phi t, \quad  q = \lambda \bar{\phi} / m_{\phi}, \quad \mathrm{and} \quad \tilde{\omega}_k = \omega_k / m_\phi,
} 
the equations can be expressed as
\dis{
    \frac{d{f}_{k}}{d \tau} &= q \left[ 2 \xi_k \sin \left( 2 \tilde{\omega}_k \tau \right)-2 \eta_{k} \cos \left(2 \tilde{\omega}_k \tau\right) \right] \cos \tau ,\\
    \frac{d{\xi}_{k}}{d \tau} &=q \left[ \left(1-2f_{k}\right) \sin (2 \tilde{\omega}_k \tau) \right] \cos \tau ,\\
    \frac{d{\eta}_k}{d\tau} &=-q \left[\left(1-2f_{k}\right) \cos \left( 2 \tilde{\omega}_k \tau \right) \right] \cos \tau.
}
where we take $m_\psi=0$, for simplicity.
We can find a relation
\dis{
(1-2f_k(t))\dot{f_k}(t) = 2\xi(t)\dot{\xi}(t) + 2\eta(t)\dot{\eta}(t) ,
}
and its integration gives
\dis{
f_k(t)(1-f_k(t)) = \xi_k^2(t) + \eta_k^2(t)= g_k(t)g_k^*(t) \geq 0,
}
where we used the initial condition $f=\xi=\eta=0$ at $t_i$.
Since $\xi_k$ and $\eta_k$ are real, the RHS of above equation should be bigger than or equal to 0. Therefore, it is easily seen that the value of $f_k(t)$ must be between 0 and 1. Additionally, the derivatives of $\xi_k$ and $\eta_k$ satisfy
\dis{
\dot{\xi}_k(t) \cos(2 \omega_k t) + \dot{\eta}_k (t) \sin (2 \omega_k t) = 0.
}

\begin{figure*}[!t]
\begin{center}
    \begin{tabular}{cc} 
     \includegraphics[width=0.45\textwidth]{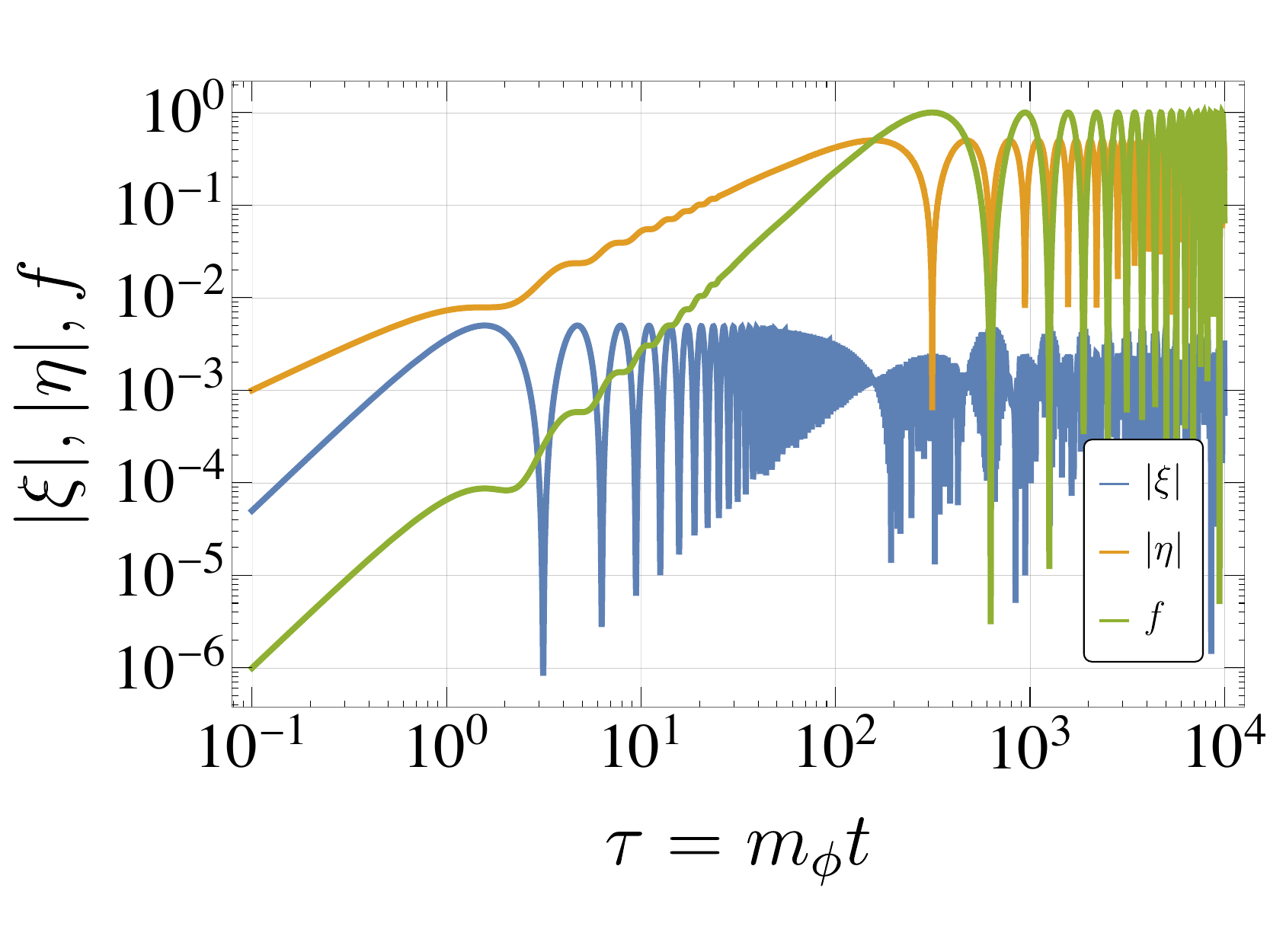}
     \includegraphics[width=0.45\textwidth]{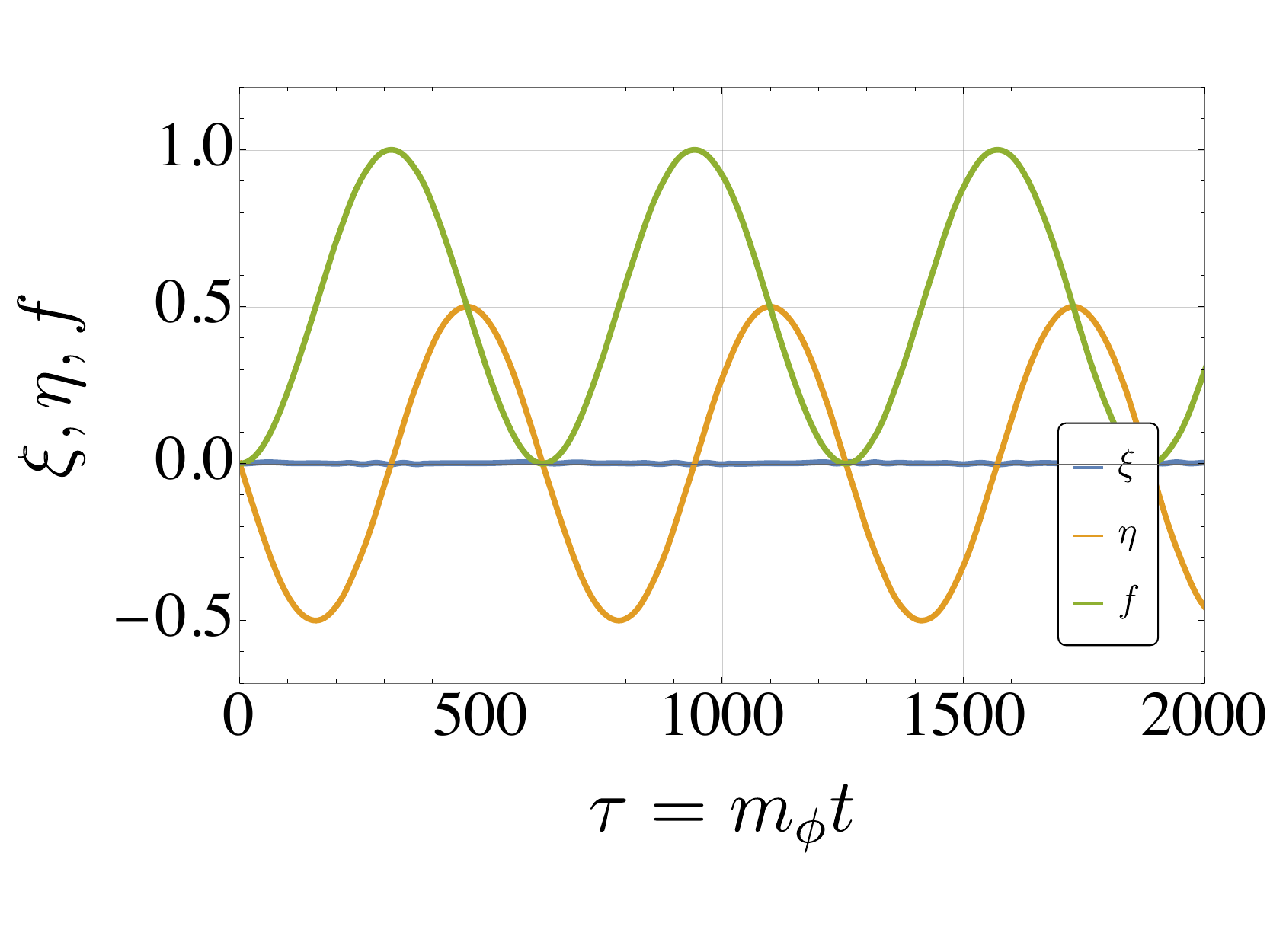}
  \end{tabular}
\end{center}
%\end{center}
\caption{Evolution of distribution function of $\psi$ by QFT calculation with $\omega = m_\phi/2$ and $q=0.01$ without considering expansion. It grows until it reaches 1 (around $\tau = \pi/q$) and starts oscillation between 0 and 1 with period $2\pi/q$. In the right figure we show the oscillation in the linear scale.}
\label{fig:qft_half}
\end{figure*}

\begin{figure*}[!t]
\begin{center}
\begin{tabular}{cc} 
 \includegraphics[width=0.45\textwidth]{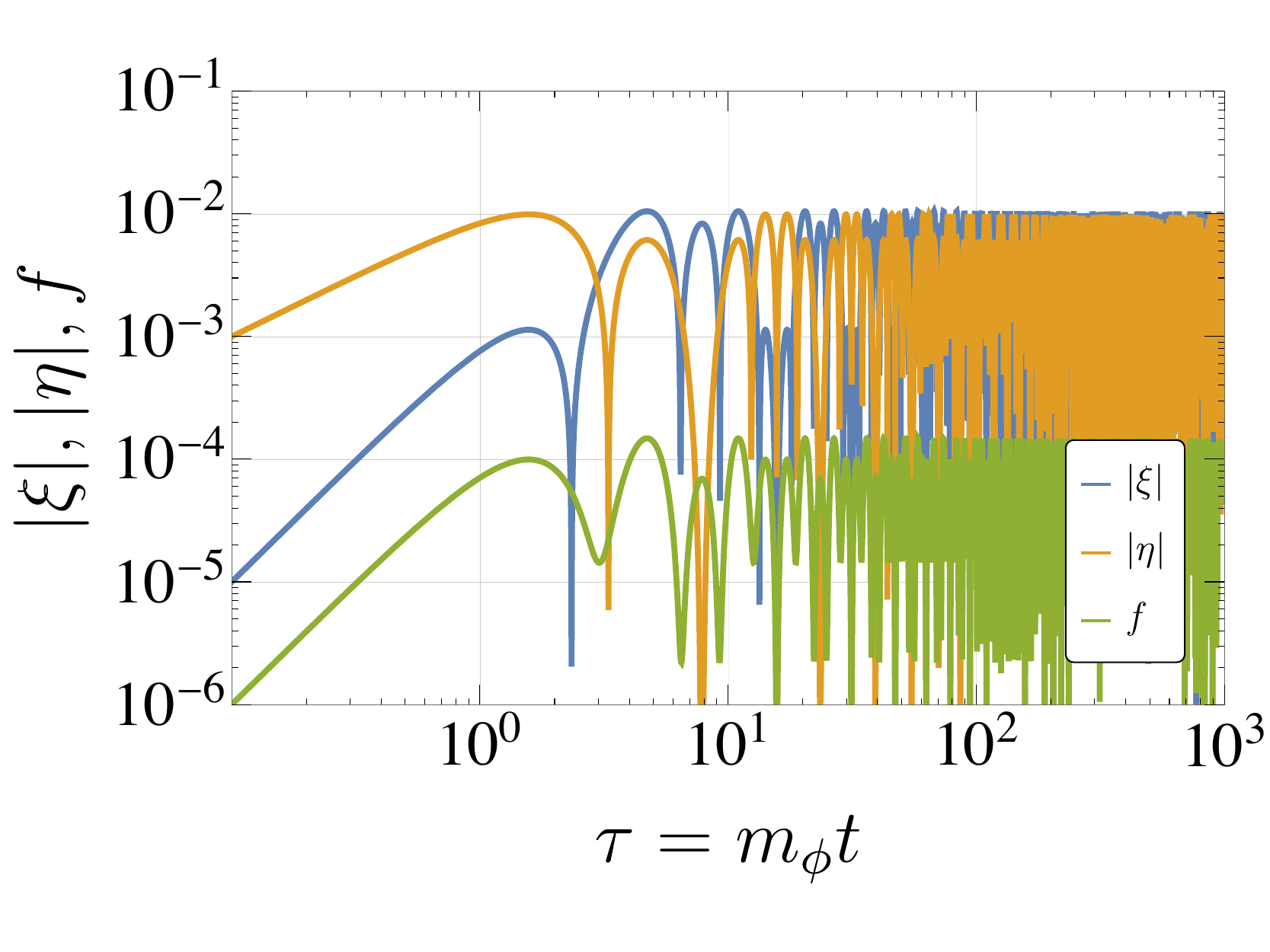}
 &
  \includegraphics[width=0.45\textwidth]
  {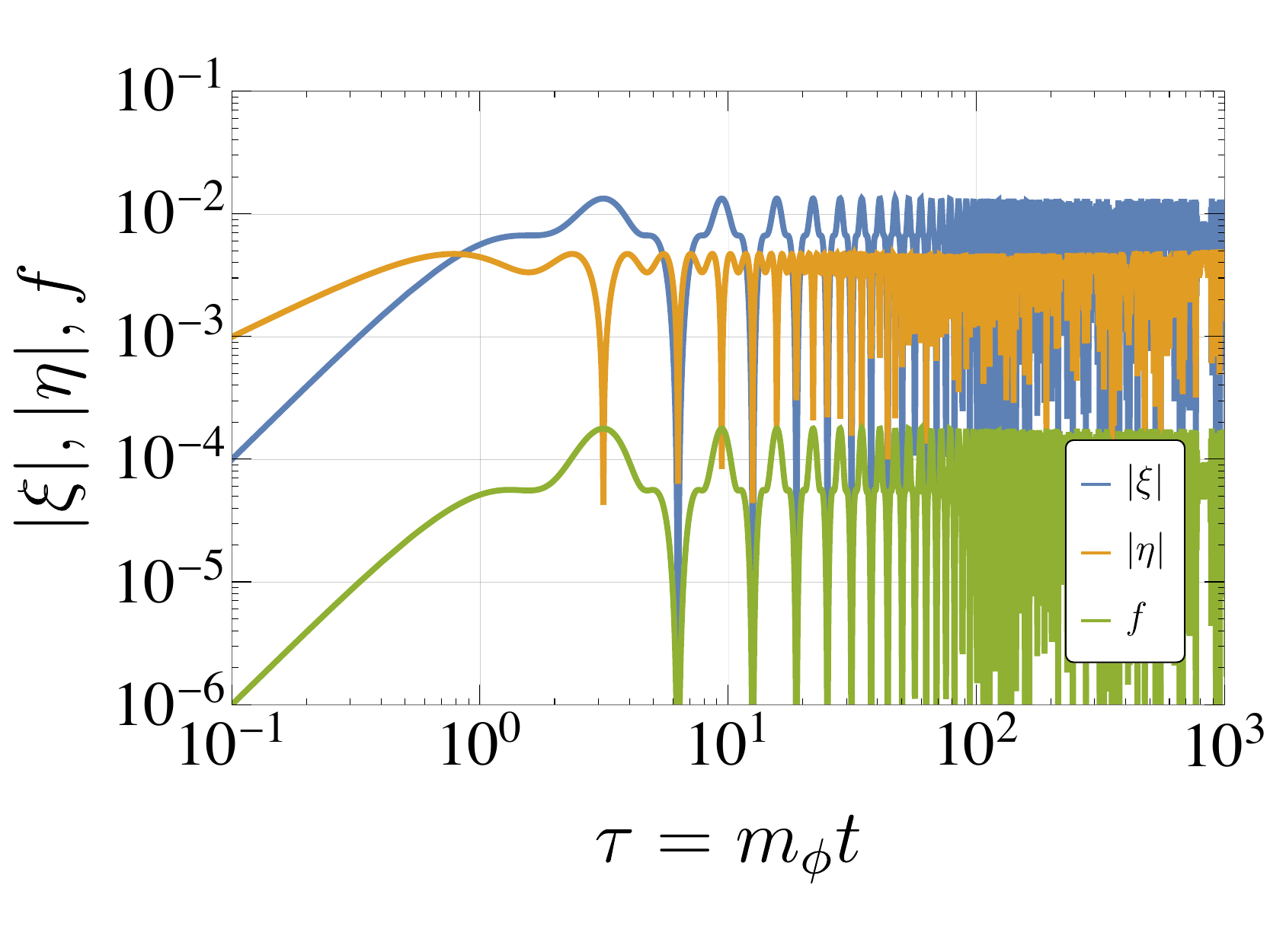}
  \end{tabular}
\end{center}
%\end{center}
\caption{Evolution of momentum distribution by QFT calculation with $\omega = 0.1m_\phi$ (left) and $\omega =  m_\phi$ (right), respectively. We used $q=0.01$ without considering expansion.}
\label{fig:qft_one}
\end{figure*}

In Fig.~\ref{fig:qft_half}, we show the numerical solution for $\omega_k = m_\phi/2$, with initial conditions $f_k(0) = \xi_k (0) = \eta_k (0) =0 $. In the left window, we can see that $f_k$ grows as a power of time until $q\tau \sim \pi$ and then oscillate between 0 and 1 with period of $2\pi/q$.
The linear plot is shown in the right window. As is shown, $\xi(t)$ is subdominant and $\eta(t)$ forces the change of $f(t)$, where $df/dt \propto -\eta(t)$.
In Fig.~\ref{fig:qft_one}, the result for $\omega_k = 0.1 m_\phi$ (left) and $m_\phi$ (right) are shown, too. In these cases, $f_k\ll1$ oscillates with small amplitude around $\sim q^2$ and cannot grow to $1$.

\begin{figure*}[!t]
\begin{center}
\begin{tabular}{cc} 
  \includegraphics[width=0.5\textwidth]{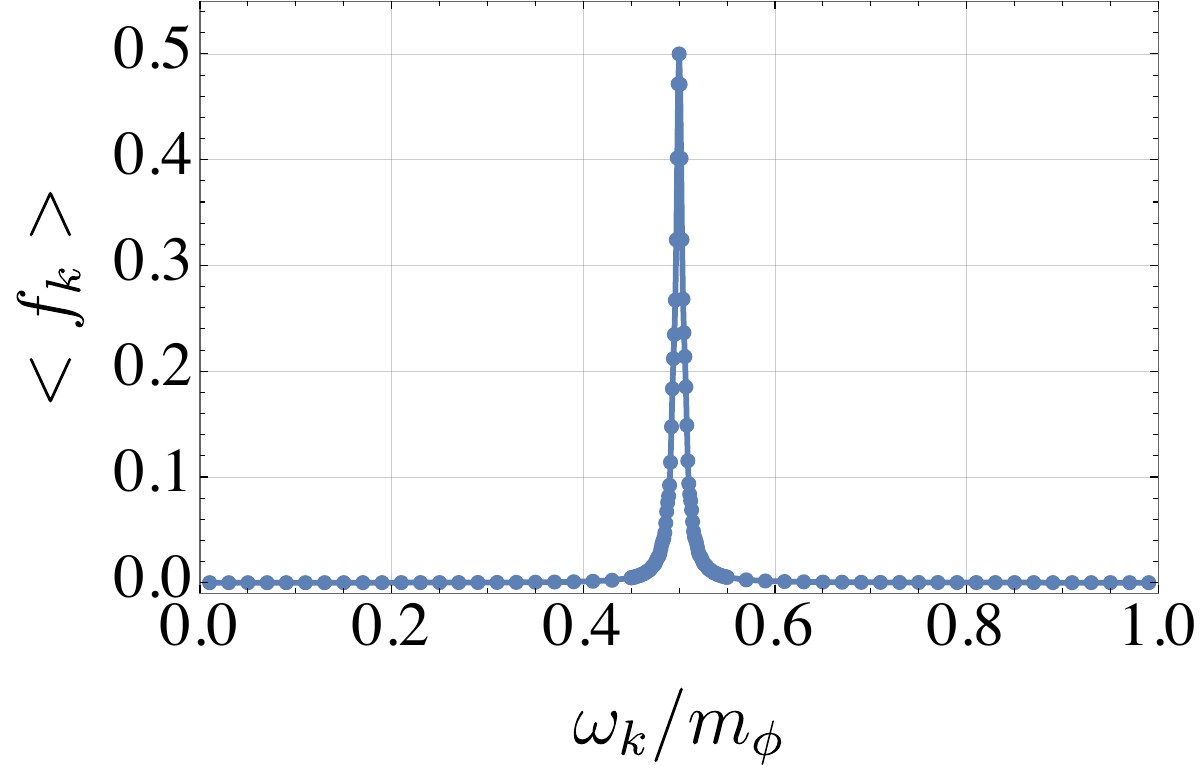}
  \end{tabular}
\end{center}
%\end{center}
\caption{The average value of $f_k$ with various $\omega_k$ without considering expansion. $\VEV{f_k}$ has a maximum value $1/2$ when $\omega_k = m_\phi / 2$. Away from $m_\phi/2$, the value of $\VEV{f_k}$ is highly suppressed.}
\label{fig:qft_average}
\end{figure*}

In Fig.~\ref{fig:qft_average}, we show the average value of $f_k$ at a late time during oscillation phase as a function of the energy of produced $\psi$ normalized by  $m_\phi$.
We can see that the large $\VEV{f_k}$ is possible only for the narrow range of $\omega_k$ around $m_\phi/2$. It has a maximum average value $\VEV{f_k}=1/2$ when $\omega_k = m_\phi / 2$, otherwise, the value of $\VEV{f_k}$ goes to 0 away from $\omega_k= m_\phi/2$.

%%%%%%%%%%%%%%%%%%%   
%%%%%%%%%%%%%%%%%%%%%

%%%%%%%%%%%%%%%%%%%   
%%%%%%%%%%%%%%%%%%%%%

%%%%%%%%%%%%%%%%%%%   
%%%%%%%%%%%%%%%%%%%%%

%%%%%%%%%%%%%
\subsubsection{Approximate Analytic  Solution}
%%%%%%%%%%%%%
We can find analytic approximations in some limiting cases. When $f_k \ll 1$, we can approximate $1-2f_k \simeq 1$, and the equations are given by
\dis{
    \frac{d{f}_{k}}{d \tau} &\simeq q \left[ 2 \xi_k \sin \left( 2 \tilde{\omega_k} \tau \right)-2 \eta_{k} \cos \left(2 \tilde{\omega_k} \tau\right) \right] \cos \tau ,\\
    \frac{d{\xi}_{k}}{d \tau} &\simeq q  \sin (2 \tilde{\omega_k} \tau)  \cos \tau ,\\
    \frac{d{\eta}_k}{d\tau} &\simeq-q  \cos \left( 2 \tilde{\omega_k} \tau \right)  \cos \tau.
}
With initial conditions $f_k(0) = \xi_k (0) = \eta_k (0) =0 $, we can find the solutions as
\begin{equation}
    \begin{split}
        f_k(\tau) &\simeq \frac{q^2 \left(1 + 12 \tilde{\omega}_k^2 + (-1 + 4 \tilde{\omega}_k^2) \cos(2 \tau) - 8 \tilde{\omega}_k (2 \tilde{\omega}_k \cos(\tau) \cos(2 \tilde{\omega}_k \tau ) + \sin(\tau) \sin(2 \tilde{\omega}_k \tau ))\right)}{2 (1 - 4 \tilde{\omega}_k^2)^2},\\
        \xi_k (\tau) &\simeq \frac{- q \left(-2 \tilde{\omega}_k + 2 \tilde{\omega}_k \cos(\tau) \cos(2 \tilde{\omega}_k \tau) + \sin(\tau) \sin(2 \tilde{\omega}_k \tau )\right)}{-1 + 4 \tilde{\omega}_k^2}, \\
        \eta_k (\tau) &\simeq \frac{q \left(\cos(2  \tilde{\omega}_k \tau) \sin(\tau) - 2 \tilde{\omega}_k \cos(\tau) \sin(2  \tilde{\omega}_k \tau)\right)}{-1 + 4 \tilde{\omega}_k^2}.
    \end{split}
    \label{eq:qft_analy_lowf}
\end{equation}
where $\tilde{\omega}_k = \omega_k/m_\phi$.
The solution of $f_k$ shows oscillating behavior with time, with the amplitude around $q^2/\tilde{\omega}_k^2$ for $\tilde{\omega}_k\gg 1/2$, or $q^2$ for $\tilde{\omega} \ll 1/2$. The value of $f_k$ is highly suppressed for small $q$ when $\omega_k$ away from 
$m_\phi/2$.

On the other hand, both the denominator and numerator in Eqs.~(\ref{eq:qft_analy_lowf}) approach zero towards $\omega_k \sim m_\phi/2$. In this limit of $\omega \rightarrow m_\phi /2$, those asymptotic form
\dis{
f_k(\tau)& \rightarrow \frac{q^2}{8} (1+2\tau^2 -\cos2\tau +2\tau \sin2\tau) \sim \frac{q^2}{4}\tau^2 ,\\
\xi_k(\tau) &\rightarrow \frac{q}{2} (1-\cos^2\tau) \sim q,\\
\eta_k(\tau)&\rightarrow -\frac{q}{2} (\tau+\cos\tau\sin\tau) \sim -\frac{q}{2}\tau,
}
show increase of $f_k$ as $f_k \sim \frac{q^2}{4}\tau^2$ for $1 \ll \tau \ll q^{-1}$, within the range $f_k\ll1$. 

The solution in \eq{eq:qft_analy_lowf} is invalid for $f_k\gtrsim 1/2$, or in other words, for the narrow resonance range
\dis{
\frac{m_\phi}{2} (1-q) \lesssim \omega_k \lesssim \frac{m_\phi}{2} (1+q).
\label{narrowband}
}
In this regime, $f_k$ cannot be ignored.
Instead, we find that the value of $\xi_k(t)$ is negligible from the numerical calculation (for example, see \ref{fig:qft_half}).
The evolution equation can be approximated as
\begin{equation}
    \begin{split}
        \frac{d{f}_{k}}{d \tau} & \simeq -2 q \eta_{k} \cos \left(2 \tilde{\omega_k} \tau\right) \cos \tau ,\\
        \frac{d{\eta}_k}{d\tau} & \simeq-q \left(1-2f_{k}\right) \cos \left( 2 \tilde{\omega_k} \tau \right) \cos \tau.
    \end{split}
\end{equation}
For $\omega_k=m_\phi/2$, the solution of  the differential equations are 
\dis{
f_k (\tau) &\simeq \sin^2\left[\frac{1}{2} q \left(\tau + \cos\tau \sin\tau\right)\right] \sim  \sin^2(q\tau/2), \\
\eta_k (\tau) &\simeq -\frac{1}{2} \sin\left[q (\tau + \cos\tau \sin\tau)\right] \sim -\frac12 \sin(q\tau).
\label{ResSol}
}
These solutions match well with the numerical solutions.
The solution $f_k(t)$ oscillates between $0$ and $1$ with the period $T(\tau) = 2 \pi / q$ for large $\tau \gg q^{-1}$. The average over time gives $\VEV{f_k(t)}=1/2$.

\subsection{QFT vs. Boltzmann equation}

Let's recall the RHS of Boltzmann equation (\ref{BE_noxpand})
\dis{
f_k^{\rm coll}(t) 
&= \Gamma_\psi n_\phi(t)\bfrac{\pi^2}{p_\psi^2}\delta(k-p_\psi)\left[ 1-2 f_k(t) \right].
\label{BERHS}
}
As shown in Ref.~\cite{Moroi:2020bkq}, for the time scale $t\lesssim (qm_\phi)^{-1}$ or small $f_k\ll 1$, the averaged time derivative of $f_k$ can be expressed as 
\dis{
\left.\VEV{\dot{f}_k}\right|_{t\lesssim (qm_\phi)^{-1}} \simeq \frac{\pi}{4} ( \lambda \bar{\phi})^2\delta(\omega-\frac{m_\phi}{2}) \simeq \frac{\pi^2 n_\phi\Gamma_\psi}{p_\psi^2}\delta(k-p_\psi),
}
which is consistent with \eq{BERHS} when $f_k\ll1$ .
This is even true for $t\gtrsim (qm_\phi)^{-1}$
outside the resonance band. 

Within the narrow band, the solution \eq{ResSol} gives a averaged value 
\dis{
\VEV{f_k} =\frac12,
}
which is consistent with the solution in \eq{BEsol1} from Boltzmann equation without cosmic expansion for  initial large $n_\phi$.
However, the time taken for growth of $f_\psi$ is \dis{
m_\phi t_c\sim q^{-1}
} 
in QFT, while it is 
\dis{
m_\phi t_c\simeq m_\phi \bfrac{n_\phi(t_i)}{p_\psi^3/(2\pi^2)}^{-1}\Gamma_\phi^{-1} \sim  \frac{9}{4\pi^2} q^{-2}, \label{tauc_BE}
}
in Boltzmann equation.

%%%%%%%%%%%%%%%%%
\subsection{With Cosmic Expansion}
\label{sec:QExp}
%%%%%%%%%%%%%%%%%
In the expanding Universe, the momentum of the fermions produced from the decay of scalar redshifts. Therefore, a specific mode which is higher than the resonance band could enter the resonance band as the Universe expands and then becomes smaller to exit the resonance band. To incorporate this redshift, we replace the time derivative in \eq{fg} as~\cite{Dodelson:2003ft}
\dis{
\frac{d}{dt} \rightarrow \frac{\partial }{\partial t} - H k \frac{\partial }{\partial k},
}
with the Hubble parameter $H$ and 
assuming the homogeneity and isotropy of the Friedmann-Robertson-Walker background. However, the RHS of \eq{fg} can be used from the flat spacetime result since the time scale of the production is much shorter than that of the Hubble expansion~\cite{Laine:2016hma}.

Taking this into account, \eq{fg} becomes,
\dis{
    \dot{f}_{k} & = k H \frac{\partial f_k}{\partial k} + i \lambda \phi (t) \left(g_{k} - g^*_{k} \right),\\
    \dot{g}_{k} & = k H \frac{\partial g_k}{\partial k} -i\left[2 \omega_k g_{k}(t)+\lambda \phi (t) \left\{ \frac{k^2}{\omega_k^2} \left(1-2f_{k}\right)+\frac{2m_\psi}{\omega_k} g_{k}\right\} \right].
    \label{fgH}}

We can simplify these equations if we use the comoving momentum as below.
With a fixed comoving momentum, $\hat{k}$, we define new functions as
\dis{
f_*(t) \equiv f_{k=K(t)} (t), \quad g_* (t) \equiv g_{k=K(t)} (t),
}
where $K(t)=\frac{a_0}{a(t)} \hat{k}$ is the physical momentum for the given comoving momentum $\hat{k}$ at a time $t$.
Here, we choose the time of $a_0$ so that the magnitude of comoving momentum $\hat{k}$ becomes the same as the physical momentum at that time.

We decompose $g_* (t)$ with real functions $\xi_*(t)$ and $\eta_*(t)$ as
\dis{
g_*(t) = e^{-2 i \Theta (t)} \left[ \xi_*(t) + i \eta_*(t) \right],
}
with $\Theta (t) = \int^t_{t_a} \Omega(t^\prime) d t^\prime$ and $\Omega(t) = \sqrt{K^2(t) + m_\psi^2}$. We can take $t_a=0$ without changing the results.
For $m_\psi=0$, $\Omega(t) = K(t)$, and
\dis{
\Theta(t) = a_0\hat{k} \int_0^t \frac{1}{a(t')}dt' = 3a_0\hat{k}t_0^{2/3}t^{1/3}.
}
In the last equality, we used the relation in the matter-dominated Universe, $a(t)=a_0(t/t_0)^{2/3}$. 

During the non-relativistic oscillating $\phi$ dominates the Universe, the solution of $\phi$ is described by 
\dis{
\phi(t) = \bar{\phi} \bfrac{a(t)}{a_0}^{-3/2} \cos(m_\phi t),
}
including the redshift due to the expansion of the Universe but ignoring the reduction due to decay. 
%By using dimensionless variables,
%\dis{
%q=\frac{\lambda \bar{\phi}}{m_\phi},\quad  \tilde{K}(t) = \frac{K(t)}{m_\phi},\quad \tilde{\Omega}(t) = \frac{\Omega(t)}{m_\phi},
%}
% and $a^3\propto \tau^2$ and taking $m_\psi=0$,
 Then, the evolution equations of the distribution functions \eq{fgH} become for comoving momentum $\hat{k}$ 
\dis{
    \frac{d{f}_{*} (\tau)}{d \tau} &= q \bfrac{\tau}{\tau_0}^{-1} \left( 2 \xi_* (\tau) \sin 2 \Theta(\tau) -2 \eta_{*} (\tau) \cos 2 \Theta(\tau) \right) \cos \tau ,\\
    \frac{d {\xi}_{*}(\tau)}{d \tau} &=q \bfrac{\tau}{\tau_0}^{-1}  \left[\left(1-2f_{*} (\tau) \right) \sin 2 \Theta(\tau) \right] \cos \tau ,\\
    \frac{d{\eta}_*(\tau)}{d \tau} &= - q \bfrac{\tau}{\tau_0}^{-1} \left[\left(1-2f_{*} (\tau)\right) \cos 2 \Theta(\tau)\right] \cos \tau,
}
with $q=\lambda \bar{\phi}/m_\phi$, and $\Theta(\tau) = 3\tilde{\omega}_k\tau_0^{2/3}\tau^{1/3}$.
Note that the equations are invariant under the scaling of $q'=\alpha q$, $\tau_0'=\alpha^{-1} \tau_0$ and $\omega_k'=\alpha^{2/3}\omega_k$.

For a given comoving momentum $\hat{k}$, the resonance occurs when its physical momentum becomes $K(t)$ becomes the half of the scalar mass, 
\begin{equation}
    K(t)=\frac{a_0}{a(t)} \hat{k} = \frac{m_\phi}{2} ,
\end{equation}
and the corresponding $\tau_k=m_\phi t_k$ is expressed as
\dis{
\tau_k\equiv m_\phi t_{K(t)=m_\phi/2}= \tau_0 \bfrac{2\hat{k}}{m_\phi}^{3/2}.
}

Now, let's consider a comoving momentum $\hat{k}$ whose physical momentum is $K(t_0)=\omega_k$ at $t_0$ (or $\tau_0=m_\phi t_0$). By using  $a_0=a(t_0)=1$ and $\tau\equiv m_\phi t$, we obtain
\dis{
\Omega(\tau)=\bfrac{\tau_0}{\tau}^{2/3}\omega_k,\qquad \Theta(\tau) = \frac{3\omega_k}{m_\phi} \tau_0\bfrac{\tau}{\tau_0}^{1/3}.
}
%At a given time $\tau$, the period of $\sin(2\Theta(\tau))$ is about \dis{
%\frac{\pi}{\tilde{w}}\bfrac{\tau_0}{\tau}^{2/3},
%}
%which can be compared to the period $\pi/ \tilde{w}_{\rm non-exp}$ in the non-expanding case with a constant $\tilde{w}_{\rm non-exp}$. Therefore, we find the correspondence of
%\dis{
%\tilde{w}_{\rm non-exp} = \tilde{w} \bfrac{\tau_0}{\tau}^{2/3}.
%}

\begin{figure*}[!t]
\begin{center}
\begin{tabular}{cc} 
 \includegraphics[width=0.6\textwidth]{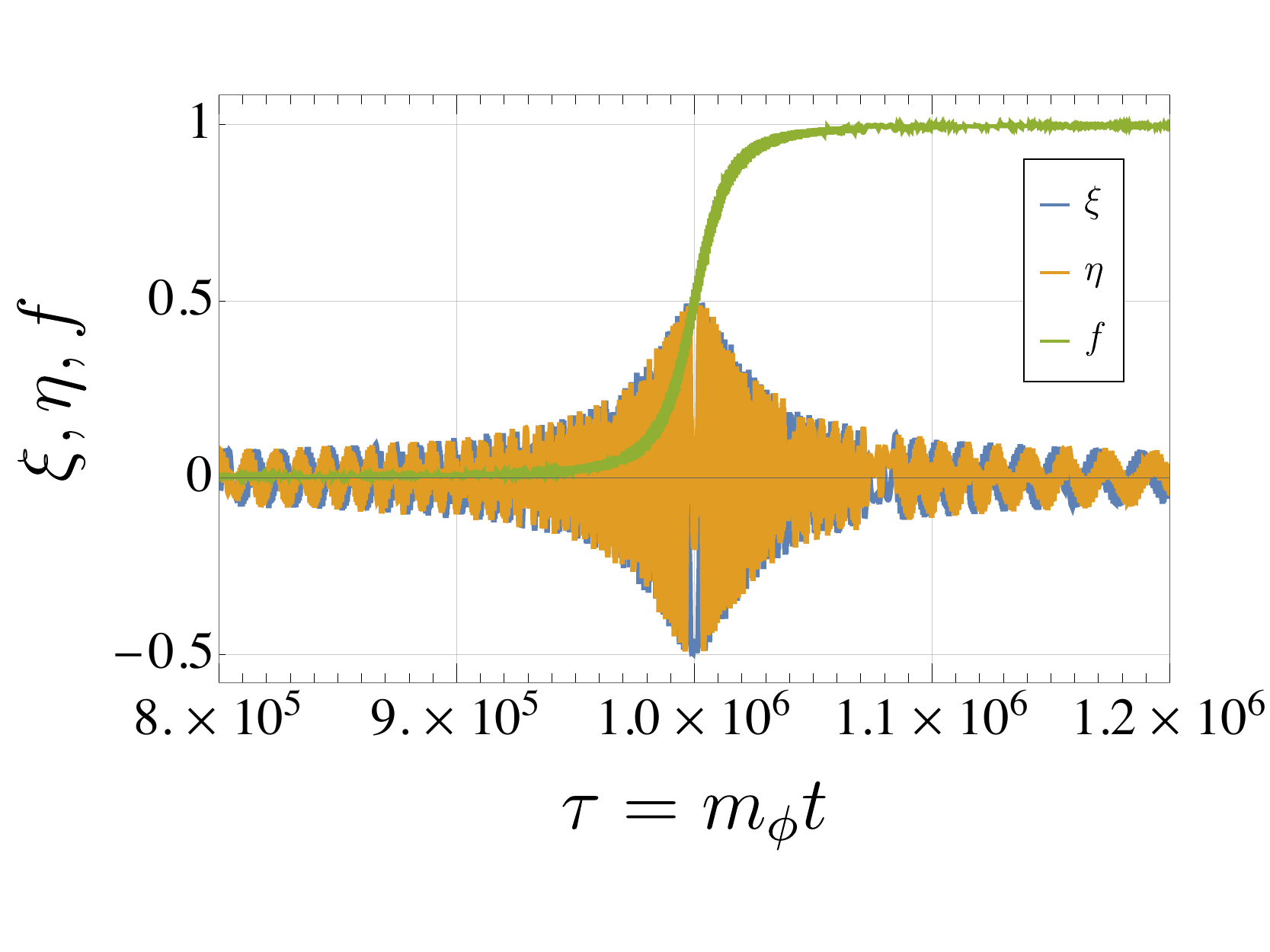}
  \end{tabular}
\end{center}
%\end{center}
\caption{Time evolution of the distribution $f_*$ as well as $\xi_*$ and $\eta_*$  for a given comoving momentum $\hat{k}$ from QFT in the expanding Universe. 
We used $q=0.01$ and $\tau_0=10^2q^{-2}$.
}
\label{fig:qft_expanding}
\end{figure*}

In Fig.~\ref{fig:qft_expanding}, we show time evolution of the distribution $f_*(t)$ as well as $\xi_*$ and $\eta_*$ for a given comoving momentum $\hat{k}$ from QFT calculation in the expanding Universe. Here, we used $q=0.01$ and $\tau_0=10^2q^{-2}$.  The rapid growth of $f_*$ happens when the corresponding physical momentum is within the narrow band in \eq{narrowband}. The corresponding time for this band becomes 
\dis{
1-\frac{2}{3}q \lesssim \frac{\tau}{\tau_0} \lesssim 1+\frac{2}{3}q.
\label{timewidth}
}
Within this range, the change of $\tau/\tau_0$ is negligible, and the term in the differential equations  can be put constant as
$q(\tau/\tau_0)^{-1} \simeq q$ for small $q\ll1$.

 When $\tau=\tau_0$, the physical momentum equals $m_\phi/2$ and the value of $f_*$ becomes $1/2$, and later grows to $1$.
This is compared to the case of Boltzmann calculation, where $f_k$ goes to $1/2$ in the expanding background. This factor 2 difference was also reported in Ref.~\cite{Moroi:2020bkq}, where for boson production
the factor appears in the exponent in the exponential function and make huge difference between BE and QFT. However, in our case, due to the Pauli blocking, it  makes only a factor 2 difference. The difference can be cured if the factor $1-2f_\psi$ in the Boltzmann equation is modified to $1-f_\psi$ as noticed in Ref.~\cite{Moroi:2020bkq}.

%%%%%%%%%%%%%%%%%%%   
%%%%%%%%%%%%%%%%%%%%%

Note that, however, as shown in the non-expanding case of the previous section, at least $\tau\gtrsim q^{-1}$ is required for growth of $f_k$ even  in the resonance band. Considering the time width $\Delta \tau\simeq q\tau_0$ in \eq{timewidth}, and requiring $\Delta \tau \gtrsim q^{-1}$, at least $\tau_0 \gg q^{-2}$ is needed for a given comoving momentum whose physics momentum becomes $m_\phi/2$ at $\tau_0$.  This is again consistent with that in Boltzmann equation, \eq{tauc_BE}.
 It means that
 the distribution function is suppressed for the comoving momentum,
 \dis{
\hat{k}\lesssim \frac{m_\phi}{2}\bfrac{1}{ q^2 \tau_0}^{2/3},
 }
 since it did not have enough time to grow.
 
In the early Universe when $q_{\rm eff}\equiv q(\tau/\tau_0)^{-1}>1$, the broad resonance may happen~\cite{Greene:1998nh,Greene:2000ew}. In this case, the fermions can be excited to the non-degenerate Fermi surface of momentum $k_{\rm eff}\sim \sqrt{q_{\rm eff}} m_\phi$. These fermions redshfit after production, and its momentum at present becomes
\dis{
k_{\rm eff}(\tau_0) = k_{\rm eff} \bfrac{\tau}{\tau_0}^{2/3} \sim \sqrt{q} \bfrac{\tau}{\tau_0}^{1/6} m_\phi \ll \frac{m_\phi}{2},
}
which is much smaller than $m_\phi/2$ at time $t_0$ once $q\ll1$. Therefore, the production in the early Universe related to the broad resonance does not affect the late time delay of DM decay.

%%%%%%%%%%%%%%
\section{A Specific Model}
\label{sec:Model}
%%%%%%%%%%%%%%%%%
We consider $\phi(t)$ as a coherently oscillating scalar dark matter with,
\dis{
\phi(t)=\bar{\phi} \bfrac{a(t)}{a(t_0)}^{-3/2} \cos(m_\phi t).
}
The relic density is
\dis{
\rho_\phi(t) = \rho_\phi(t_0) \bfrac{a(t)}{a(t_0)}^{-3} \simeq  \rho_\phi(t_0) \bfrac{t}{t_0}^{-2},
}
where $t_0$ is the present age of the Universe and we assumed DM domination neglecting the dark energy component, for simplicity.
The present energy density of DM is%~\footnote{In our paper, we focused on the 100\% DM made of $\phi$ and extended lifetime to be larger than the age of the Universe. However, the same mechanism can be applied to the subdominant component dark matter which may decay earlier than today in the models to explain cosmological problems~\cite{Berezhiani:2015yta,Chudaykin:2016yfk}.}
\dis{
\rho_\phi(t_0)=\frac12 m_\phi^2 \bar{\phi}^2\simeq 3.7\times 10^{-47}\gev^4,
\label{DMdensity}
}
and the number density is
\dis{
n_\phi(t) = \frac{\rho_\phi(t)}{m_\phi}=3.7\times 10^{-35}\gev^3 \bfrac{10^{-3}\unitev}{m_\phi}\bfrac{t}{t_0}^{-2}.
\label{nphit0}
}
For a fixed DM energy density in \eq{DMdensity}, the variable $q$  can be written as
\dis{
q\equiv\frac{\lambda \bar{\phi}}{m_\phi} = 8.6 \times 10^{-12} \bfrac{\lambda}{10^{-12}}\bfrac{10^{-3}\unitev}{m_\phi}^2.
}
The lifetime without Pauli blocking in~\eq{lifetime0} is
\dis{
\tau_{\phi,0} \simeq 1.6 \times 10^{13} \sec \bfrac{10^{-12}}{\lambda}^2\bfrac{10^{-3}\unitev}{m_\phi},
}
however this lifetime is extended due to the Pauli blocking. 

As explained in  \eq{nphi_t}, the scalar $\phi$ would decay quickly when the number density becomes smaller than that in the Fermi sphere, $n_\phi(t) \lesssim \frac{p_\psi^3}{6\pi^2} $, when the Pauli blocking is inefficient. Using this equation and \eq{nphit0} with $p_\psi \simeq m_\phi/2$, we find the extended lifetime as
\dis{
\tau_{\phi} \simeq 132 \times t_0 \bfrac{10^{-3}\unitev}{m_\phi}^2,
}
which is independent of the coupling $\lambda$, once $\tau_{\phi}>\tau_{\phi,0}$.

There is also a constraint on the mass of $\phi$, $m_\phi$, for Pauli blocking to work. To fill the Fermi sphere with Fermi momentum $p_\psi$, we need at least a number density larger than the number density in the Fermi sphere as
    \dis{
    n_\phi \gg \frac{p_\psi^3}{6\pi^2}.
    }
Using \eq{DMdensity} and \eq{nphit0} with $p_\psi \simeq m_\phi/2$ for $m_\psi\ll m_\phi$, we find the upper bound on the mass
    \dis{
    m_\phi \lesssim (16\pi^2 \rho_{\rm DM})^{1/4} \simeq 0.01\, {\rm eV}.
    }
    Therefore for larger value of $m_\phi$ than this, the Pauli blocking does not work due to the insufficient number density of $\phi$. Note that, for subdominant DM component, the constraint becomes stronger since its energy density is smaller by the fractional factor.

In this paper, we consider that $\phi$ and $\psi$ do not have further scattering between each other except the decay and inverse decay, since the scatterings may redistribute the momentum spectrum of $\psi$ and destroy the Pauli blocking. For this, we need small Yukawa coupling, so that  the scattering rate $n\langle \sigma v \rangle \sim  \lambda^4 \rho_\phi/m_\phi^3$ is much smaller than the Hubble rate $H$ at present. This gives the upper bound on the Yukawa coupling as 
    \dis{
    \lambda \ll \left( \frac{H m_\phi^3}{\rho_\phi}\right)^{1/4} \simeq 2\times 10^{-8}\bfrac{m_\phi}{\rm 10^{-3}eV}^{3/4}.
    }
    For this parameter, the Pauli blocking is not affected in our model.
    
    For very small Yukawa coupling and mass, the lifetime in the vacuum becomes longer than the age of the Universe, then the Pauli blocking becomes meaningless at the present Universe.

The number density and energy density of neutrinos are constant as
\dis{
n_\psi = \frac{p_\psi^3}{6\pi^2},\qquad \rho_\psi = \frac{m_\phi^4}{128\pi^2}.
}
From the conservation of energy in the expanding Universe, it is required that
\dis{
\frac{d}{dt}(\rho_\phi + \rho_\psi) + 3H\rho_\phi +4H\rho_\psi=0,
}
where the dominant neutrinos are relativistic. Therefore, the evolution of the energy density of $\phi$ is found to be~\cite{Bjaelde:2010vt}
\dis{
\rho_\phi(t) = -\frac{m_\phi^4}{96\pi^2} + \left(\rho_\phi(t_i) + \frac{m_\phi^4}{96\pi^2}\right)\bfrac{a(t)}{a(t_i)}^{-3},
}
for $t\gg t_i$.

As a specific example, we propose a realistic particle physics model where the light scalar is dark matter and its long life time is realized because the decay into the SM neutrinos is blocked by the cosmic neutrino background. 
We extend the SM to address the non-vanishing mass of neutrinos with $SU(2)$ triplet Higgs fields, so-called type-II seesaw mechanism~\cite{Schechter:1980gr,Magg:1980ut,Cheng:1980qt}.  
The Yukawa couplings between the left-handed lepton doublet $l_L$ and the triplet Higgs field $\Delta$ is given by
\begin{equation}
    \mathcal{L}_{\text {Yukawa }} \supset-\frac{1}{\sqrt{2}} Y_{\Delta}^{i j} \overline{l_{L}^{i C}} \cdot \Delta l_{L}^{j}+ h.c.
    \label{eq:tripletYukawa}
\end{equation}
After the triplet Higgs field develops the vacuum expectation value (VEV) $v_{\Delta}$, left-handed neutrino masses are generated as
\begin{equation}
    \left(\mathcal{M}_{\nu}\right)_{i j}=Y_{\Delta}^{i j} v_{\Delta}.
\end{equation}
We note that the upper bound $v_{\Delta} \lesssim 0.78$~GeV is obtained~\cite{Okada:2022cby} by interpreting the so-called $\rho$ parameter constraint~\cite{ParticleDataGroup:2022pth}\footnote{The lower bound on $v_{\Delta} \gtrsim 1$~eV is also obtained~\cite{Antusch:2018svb}, however this depends on the extra assumption of the charged Higgs boson mass.}. 
The scalar potential is given by
\begin{align}
    V=     & V_{1}+V_{2}  \\
    V_{1}= & -\mu_{1}{ }^{2}\left|\Phi\right|^{2}+\frac{1}{2} \lambda_{1}\left|\Phi\right|^{4} \nonumber \\
      & +\mu_{3}{ }^{2} \operatorname{Tr}\left(\Delta^{\dagger} \Delta\right)+\frac{1}{2} \Lambda_{1}\left(\operatorname{Tr}\left(\Delta^{\dagger} \Delta\right)\right)^{2}+\frac{1}{2} \Lambda_{2}\left(\left(\operatorname{Tr}\left(\Delta^{\dagger} \Delta\right)\right)^{2}-\operatorname{Tr}\left[\left(\Delta^{\dagger} \Delta\right)^{2}\right]\right) \nonumber \\
      & +\Lambda_{4}\left|\Phi\right|^{2} \operatorname{Tr}\left(\Delta^{\dagger} \Delta\right)+\Lambda_{5} \Phi^{\dagger}\left[\Delta^{\dagger}, \Delta\right] \Phi-\frac{\Lambda_{6}}{\sqrt{2}}\left(\Phi^{T} \cdot \Delta \Phi+\text { H.c. }\right)   \\
    V_{2}= & \frac{1}{2}\mu_{S}^{2} S^{2}+\frac{1}{4} \lambda_{S} S^{4}+\frac{1}{2} \lambda_{S 1} S^{2}\left|\Phi\right|^{2}+\frac{1}{2} \lambda_{S \Delta} S^{2} \operatorname{Tr}\left(\Delta^{\dagger} \Delta\right)-\frac{\lambda_{6}}{\sqrt{2}} S\left(\Phi^{T} \cdot \Delta \Phi+\text { H.c. }\right)
\end{align}
where $\Phi$ is the SM doublet Higgs and $S$ is the real singlet scalar of dark matter candidate. 
Around the VEVs of $\Delta$, $\Phi$ and $S$, which are $v_{\Delta}, v\simeq 246$ GeV and $v_s$ respectively, the real neutral components of those fields are expanded as
\begin{align}
& \Delta = \left(
\begin{array}{cc}
0 &, 0 \\
 \frac{1}{\sqrt{2}}(v_\Delta+\delta^0)& , 0
\end{array}
\right) ,\\
& \Phi = \left(
\begin{array}{c}
0 \\
\frac{1}{\sqrt{2}}(v+\phi^0) 
\end{array}
\right), \\
& S = v_s+s  ,
\end{align}
and VEVs are expressed as
\begin{align}
 v_{\Delta} &\simeq -\frac{\lambda_{S1} \Lambda_6 v^2}{v^2 \left(\lambda_{S1} (\Lambda_4-\Lambda_5)-2 \lambda_6^2\right)+2 \lambda_{S1} \mu_3^2}, \\
 v_s &\simeq  -\frac{\lambda_6}{\lambda_{S1}}v_{\Delta}.
\end{align}
%

%The neutral component of $S, \Phi$ and $\Delta$ are denoted by $s, \phi^{0}$ and $\delta^{0}$, respectively. 
The mass squared matrix
\begin{equation}
    M_{i j}^{2}, \quad \mathrm{where} \quad(i, j) \quad \mathrm{ run } \quad \left(s, \phi^{0}, \delta^{0}\right)
\end{equation}
is diagonalized for mass eigenstates $\phi_{1}, \phi_{2}$, and $\phi_{3}$ with $m_{\phi_{1}} \ll m_{\phi_{2}} \simeq 125 \mathrm{GeV} \ll m_{\phi_{3}}$ by the unitary matrix
\begin{equation}
    U=\left(\begin{array}{ccc}
        U_{s 1}          & U_{s 2}      & U_{s 3}          \\
        U_{\phi^{0} 1}   & U_{\phi^{0} 2}   & U_{\phi^{0} 3}   \\
        U_{\delta^{0} 1} & U_{\delta^{0} 2} & U_{\delta^{0} 3}
    \end{array}\right) .
\end{equation}

The DM $\phi_{1}$ decays into two neutrino pair through the mixing $U_{\delta^{0} 1}$ and the Yukawa coupling (\ref{eq:tripletYukawa}). The decay rate is
\begin{align}
    \Gamma\left(\phi_{1} \rightarrow \nu \nu\right) & \sim \frac{1}{4 \pi}\left|Y_{\Delta} U_{\delta^{0} 1}\right|^{2} m_{\phi_{1}}  \nonumber \\
    & \sim \frac{m_{\phi_{1}}}{4 \pi} \frac{m_{\nu}^{2}}{v_{\Delta}^{2}}\left|U_{\delta^{0} 1}\right|^{2}  \nonumber \\
    & \simeq 8 \times 10^{-43} \gev \bfrac{m_{\phi_1}}{10^{-3} \unitev} \bfrac{m_\nu}{10^{-6} \unitev}^2 \bfrac{1\kev}{v_\Delta}^2 \bfrac{\abs{U_{\delta^{0} 1}}^2}{10^{-11}}
%    &\sim\frac{m_{\phi_{1}}}{4 \pi} \frac{m_\nu^2}{v_s^2}
    \label{eq:GammaPhi1},
\end{align}
%
%\dis{
%\lambda =\frac{m_\nu}{v_s} = 10^{-12}
%}
with
\begin{align}
 U_{\delta^{0} 1} \sim & \frac{M_{s \delta^{0}}^{2}}{M_{\delta^{0} \delta^{0}}^{2}} \simeq \frac{\lambda_{6} v^{2}}{{\mu}_{3}^{2}} \simeq 10^{-5}\left(\frac{\lambda_{6}}{0.1} \right)\left(\frac{10^{4} \mathrm{GeV}}{\mu_{3}}\right)^{2}. \label{eq:Udelta1}
\end{align}
where we have assumed $\mu_3^2 \gg v^2 $.
Though we normalized the decay rate $\Gamma$ in Eq.~(\ref{eq:GammaPhi1}) by about the age of Universe $t_0$ and the actual bound is $\Gamma^{-1} > \mathcal{O}(10) t_0$~\cite{Ichiki:2004vi,Audren:2014bca,Nygaard:2020sow}, as we have seen above that smaller $t_{\phi}$ is enough to have long enough $t_{\phi,ext}$. In fact, a much larger element (\ref{eq:Udelta1}) is allowed.
Since the neutrinos are in the weak $SU(2)$ doublet, in fact, DM $\phi$ can decay into two photons through loop processes involving neutrinos, the $W$ boson and electrons. 
This decay rate is very suppressed to be small enough~\cite{Heeck:2019guh}.

The neutrinos produced from the decay of DM occupies the Fermi surface inside the phase space with the momentum of $p_\psi=m_{\phi_{1}}/2$. The number density of this neutrino corresponds to 
\dis{
n_\nu = \frac{p_\psi^3}{6\pi^2} = 2.7\times 10^5 {\rm cm}^{-3}\bfrac{m_\phi}{0.01 \unitev}^3.
}
This is comparable to the cosmic background neutrino in the standard cosmological model which is
\dis{
n_{\nu,\rm SM} = 56 {\rm cm}^{-3},
}
per flavor.
The enhanced number density might be probed in the future experiments of the cosmic neutrino background like PTOLEMY-like experiment~\cite{PTOLEMY:2019hkd}.

%%%%%%%%%%%%%%
\section{Conclusion}
\label{sec:con}
%%%%%%%%%%%%%%
We studied a mechanism to make dark matter stable based on the Pauli blocking in the fermion background. We derived the evolution equations from  quantum field theory and compared them with usual Boltzmann equations. They are consistent each other, however there is a factor 2 in the expanding Universe.

When the scalar field can decay only to the fermions, the  Pauli blocking can delay the decay of the DM and lengthen the lifetime getting longer than the age of the Universe. As a realistic example, we proposed a particle physics model with a light scalar which can decay into the lightest neutrino and the Pauli blocking make its lifetime longer. In this model, we can find new parameter space becomes compatible with the observations.\\

Note added: While preparing the manuscript, we found a paper by Batell 
and Yin~\cite{Batell:2024hzo}, which has some overlap with our analyses based on Boltzmann 
equation formalism.\\

\section*{Acknowledgements}
W.Cho, K.-Y. Choi, and J.Joh acknowledge the financial support from National Research Foundation(NRF) grant
funded by the Korea government (MEST) NRF-2022R1A2C1005050.
This work of OS was supported in part by JSPS KAKENHI Grant Number 23K03402.

\appendix

\bibliographystyle{elsarticle-num}
\bibliography{reference}
% Produces the bibliography via BibTeX.
%\bibliographystyle{plain}
%\nocite{*}

\end{document}